\documentclass[twocolumn]{aastex63}

\usepackage[version=4]{mhchem}
\received{}
\revised{}
\accepted{}
\submitjournal{ApJ}

\shorttitle{chemical desorption of H$_2$S and PH$_3$}
\shortauthors{Furuya et al.}

\newcommand{\react}[4]{{\rm \ce{#1}} + {\rm \ce{#2}} \rightarrow {\rm \ce{#3}} + {\rm \ce{#4}}}

\newcommand{\reacti}[3]{{\rm \ce{#1}} + {\rm \ce{#2}} \rightarrow {\rm \ce{#3}}}

\newcommand{\edesthresh}{{E^{\rm thresh}_{\rm  b}}}
\newcommand{\edesmin}{{E^{\rm min}_{\rm  b}}}
\newcommand{\edes}{{E_{\rm b}}}
\newcommand{\ehop}{{E_{\rm hop}}}
\newcommand{\pcd}{P_{\rm cd}}
\newcommand{\pcds}{P_{\rm cd}({\rm H_2S})}
\newcommand{\pcdp}{P_{\rm cd}({\rm PH_3})}

\begin{document}

\title{Quantifying the chemical desorption of H$_2$S and PH$_3$ from amorphous water ice surfaces}

\correspondingauthor{Kenji Furuya}
\email{kenji.furuya@nao.ac.jp}

\author[0000-0002-2026-8157]{Kenji Furuya}
\affiliation{National Astronomical Observatory of Japan, Osawa 2-21-1, Mitaka, Tokyo 181-8588, Japan}
\author[0000-0002-6852-3604]{Yasuhiro Oba}
\affiliation{Institute of Low Temperature Science, Hokkaido University, Sapporo, Hokkaido 060-0819, Japan}
\author[0000-0002-0095-3624]{Takashi Shimonishi}
\affiliation{Center for Transdisciplinary Research, Niigata University, Ikarashi-ninocho 8050, Nishi-ku, Niigata, 950-2181, Japan}
\affiliation{Environmental Science Program, Department of Science, Faculty of Science, Niigata University, Ikarashi-ninocho 8050, Nishi-ku, Niigata, 950-2181, Japan}



\begin{abstract}
Nonthermal desorption of molecules from icy grain surfaces is required to explain molecular line observations in the cold gas of star-forming regions.
Chemical desorption is one of the nonthermal desorption processes and is driven by the energy released by chemical reactions. 
After an exothermic surface reaction, the excess energy is transferred to products' translational energy in the direction perpendicular to the surface, leading to desorption.
The desorption probability of product species, especially that of product species from water ice surfaces, is not well understood.
This uncertainty limits our understanding of the interplay between gas-phase and ice surface chemistry.
In the present work, we constrain the desorption probability of \ce{H2S} and \ce{PH3} per reaction event on porous amorphous solid water (ASW) by numerically simulating previous laboratory experiments.
Adopting the microscopic kinetic Monte Carlo method, we find that the desorption probabilities of \ce{H2S} and \ce{PH3} from porous ASW per hydrogen addition event of the precursor species are $3 \pm 1.5$\% and $4 \pm 2$\%, respectively.
These probabilities are consistent with a theoretical model of chemical desorption proposed in the literature if $\sim$7\% of energy released by the reactions is transferred to the translational excitation of the products.
As a byproduct, we find that approximately 70\% (40\%) of adsorption sites for atomic H on porous ASW should have a binding energy lower than $\sim$300 K ($\sim$200 K).
The astrochemical implications of our findings are briefly discussed.
\end{abstract}

\keywords{Astrochemistry(75) --- Reaction catalysts(2080)  --- Interstellar molecules(849) --- Interstellar dust processes(838)}




\section{Introduction}
\label{sec:intro}
Various molecules have been detected in the cold ($\sim$10 K) gas of star-forming regions \citep[see e.g.,][]{mcguire18}.
Some molecules, such as CO and \ce{N2H+}, have gas-phase production pathways that are sufficiently efficient to explain their observed abundances, whereas others, such as \ce{CH3OH} and \ce{H2S}, do not form by gas-phase reactions efficiently enough to explain the observations \citep[e.g.,][]{geppert06,garrod07}.
Molecules of the latter type are thought to form on dust grains via surface reactions, and to be released into the gas phase via nonthermal desorption processes because thermal desorption is negligible at 10 K.
Thus far, several mechanisms of nonthermal desorption have been proposed, including UV photodesorption \citep[e.g.,][]{hama09,bertin16,fuente17}, stochastic heating and/or sputtering by cosmic rays \citep[e.g.,][]{dartois15,ivlev15}, and chemical desorption \citep[e.g.,][]{dulieu13,minissale16,he17,chuang18,oba18,nguyen20}.
A quantitative understanding of nonthermal desorption processes is critical for understanding the interplay between gas-phase and ice surface chemistry.

Among nonthermal desorption processes, chemical desorption is the topic of the present paper.
Chemical desorption is caused by the energy released by chemical reactions; after an exothermic surface reaction, (the part of) excess energy goes into the products' translational energy in the direction perpendicular to the surface, leading to desorption \citep[e.g.,][]{fredon17}. 
Laboratory experiments have demonstrated that chemical desorption indeed occurs in several reaction systems on astrophysically relevant surfaces \citep[e.g.,][]{dulieu13,minissale16,he17,chuang18,oba18}.
In most of these studies, chemical desorption was quantified solely by gas-phase measurements using quadrupole mass spectroscopy (QMS) even though chemical desorption is a surface process.
One important result obtained from these experiments is that the fraction of reaction products desorbed from the surface depends on both the reaction and the surface composition.
The fraction on amorphous solid water (ASW) is lower than that on graphite or silicate surfaces \citep{minissale16}.
For some of the reactions studied thus far (e.g., O + O $\rightarrow$ \ce{O2}), the fraction of reaction products released from the surface is reported to be several tens of percent on graphite/silicate surfaces; by contrast, on ASW, it falls below the upper  limit of $\sim$10\% measurable in the experiments \citep[see Table 1 in ][for summary]{minissale16}.
Quantitative understanding of the chemical desorption process, in particular, on ASW is limited because of its low efficiency, although ASW would be more representative of the surface of interstellar grains in star-forming regions than graphite/silicate. 
Because of our limited understanding of chemical desorption, current astrochemical models often assume that the desorption probability per reactive event is $\sim$1\% \citep[e.g.,][]{garrod13,taquet14,furuya15}. 

For better understanding of chemical desorption on ASW, researchers have conducted experimental studies in which time-resolved infrared measurements were used to monitor surface species \citep[][see also \citet{chuang18} 
for chemical desorption due to reactions involving CO and hydrogen]{oba18,oba19,nguyen20,nguyen21}.
\citet{oba18, oba19} studied chemical desorption upon the reaction of H$_2$S with H atoms on ASW.
They deposited atomic H onto ASW partly covered by \ce{H2S}.
The following reactions can occur in this system:
\begin{align}
&\react{H2S}{H}{HS}{H2}, \label{react:h2s} \\
&\reacti{HS}{H}{H2S}. \label{react:hs}
\end{align}
Reactions \ref{react:h2s} and \ref{react:hs} are exothermic, with exothermicities of 58 kJ mol$^{-1}$ (corresponding to $\sim$7000 K) and 374 kJ mol$^{-1}$ ($\sim$45,000 K), respectively \citep{oba18}, 
which are much larger than the binding energies of \ce{H2S} and HS on ASW \citep[2700 K;][]{collings04,wakelam17}.
HS is formed from \ce{H2S} via a hydrogen abstraction reaction, and hydrogenation of HS leads to the (re)formation of \ce{H2S}.
\ce{H2S} and HS can desorb by chemical desorption in each cycle of this \ce{H2S}--HS loop. 
Because the \ce{H2S}--HS loop can continue as long as H atoms are available, a substantial fraction of \ce{H2S} and HS can desorb from the ASW surface during H-atom deposition, even if the probability of chemical desorption per reactive event is small.
In \citet{oba19}, during atomic H deposition experiments, the authors monitored the abundance of \ce{H2S} on ASW by Fourier transform infrared (FTIR) spectroscopy.
During H-atom deposition, a decrease in the intensity of the band at 2570 cm$^{-1}$, which was assigned to the S--H stretching band of \ce{H2S}, was observed.
Reaction \ref{react:h2s} has an activation energy barrier of 1560 K \citep{lamberts17}, whereas Reaction \ref{react:hs} is barrierless.
Most S on ASW might exist as \ce{H2S} during the \ce{H2S}--HS loop (this is discussed in Section \ref{sec:result}).
If so, the decrease in the intensity of the band at 2570 cm$^{-1}$ should be due to chemical desorption of \ce{H2S} and/or HS.
They found that the amount of \ce{H2S} on porous ASW gradually decreased with time (i.e., with increasing amount of H-atom deposition), that $\sim$60\% of the initial amount of \ce{H2S} was eventually lost from the ASW surface in experiments at 10 K and 20 K, and that $\sim$30\% was lost in an experiment at 30 K.
Although the dominant chemical form of S released from the surface (either \ce{H2S} or HS) could not be identified,
\ce{H2S} could be the dominant form because the exothermicity of Reaction \ref{react:hs} is greater than that of Reaction \ref{react:h2s} \citep{oba18}.

\citet{nguyen20,nguyen21} experimentally studied the chemical desorption process upon reaction of PH$_3$ with H atoms on ASW in experiments similar to those of \citet{oba18,oba19}.
They deposited atomic H onto ASW, which was partly covered by \ce{PH3}.
The following reactions can occur in this system:
\begin{align}
&\react{PH3}{H}{PH2}{H2}, \label{react:ph3} \\
&\reacti{PH2}{H}{PH3}. \label{react:ph2}
\end{align}
Reactions \ref{react:ph3} and \ref{react:ph2} are exothermic, with exothermicities of 81.5 kJ mol$^{-1}$ and 337.5 kJ mol$^{-1}$, respectively \citep{molpeceres21}.
Similar to the \ce{H2S}--HS loop, interconversion between \ce{PH3} and \ce{PH2} occurs through Reactions \ref{react:ph3} and \ref{react:ph2}. In their experiments at 10 K, $\sim$80\% of the initial amount of \ce{PH3} deposited onto porous ASW was released from the surface. 

\citet{oba19} and \citet{nguyen21} estimated a chemical desorption probability {\it per reactive species} (i.e., per incident H atoms) of $\sim$1\% on porous ASW for the \ce{H2S} + H and \ce{PH3} + H systems.
As noted by \citet{oba18}, the desorption probability per incident H atom corresponds to the lower limit of the desorption probability {\it per reactive event}, which astrochemical models require as inputs.
This is because H atoms adsorbed onto ASW surfaces can be thermally desorbed or consumed by the \ce{H2} formation via the recombination of two H atoms on the surface. 
The primary goal of this paper is to constrain the desorption probability of \ce{H2S} and \ce{PH3} per reaction event on porous ASW on the basis of the experimental data reported by \citet{oba19} and \citet{nguyen21}, respectively.
To this end, we conduct kinetic Monte Carlo simulations of their experiments.

The rest of this paper is organized as follows: 
our numerical model is described in Section \ref{sec:model}, and the results are described in Section \ref{sec:result}.
We compare our derived desorption probability with the theoretical predictions reported in the literature in Section \ref{sec:discuss}.
The astrochemical implications of the chemical desorption of \ce{H2S} and \ce{PH3} are briefly discussed.
We summarize our findings in Section \ref{sec:concl}.

\section{Methods}
\label{sec:model}
We intend to constrain the chemical desorption probability per reaction event ($\pcd$) for \ce{H2S} + H and \ce{PH3} + H systems on porous ASW by numerically simulating the experiments of \citet{oba19} and \citet{nguyen21}.
For this purpose, we adopt an on-lattice kinetic Monte Carlo (kMC) method \citep{gillespie76,cuppen13}.
In our kMC model, a sequence of processes---adsorption, thermal hopping, and desorption---are modeled as a Markov chain adopting the next reaction method \citep[e.g.,][]{gibson00,chang12}.
In the case of thermal hopping, a hop in a different direction is treated as a distinct event.
This sequence is chosen on the basis of random numbers in combination with the rates for these processes.
In contrast to the rate-equation method, which is widely used in the astrochemical community, the position and movement of each chemical species on surfaces are tracked over time in our kMC simulations.
Consideration of the binding energy distribution on surfaces in the kMC simulations is then straightforward.
Water ice surfaces are known to contain various adsorption sites with different energy depths \citep[e.g.,][]{amiaud06,hama12,karssemeijer14}.
We will show that the binding energy distribution of atomic H is a key parameter for reproducing the experiments (see Section \ref{sec:result}).
Details of the numerical aspects of the kMC method can be found in \citet{chang12}, \citet{cuppen13}, and references therein.

\citet{oba19} experimentally investigated the surface reactions of \ce{H2S} with H atoms on porous ASW, nonporous ASW, and polycrystalline water ice.
They found no strong correlation between the ice structure and the desorption probability per incident H atom in their experiments at 10 K.
In the present work, we focus on experiments on porous ASW because the experiments were conducted at three different surface temperatures: 10 K, 20 K, and 30 K.
For nonporous ASW and polycrystalline water ice, the experiments were conducted only at 10 K.
As shown later, we need experimental data at different surface temperatures to constrain the $\pcd$ and resolve the degeneracy between parameters---in particular, the binding energy distribution for atomic H and the $\pcd$.
Notably, the $\pcd$ is assumed to be independent of the surface temperature throughout this work.
This assumption would be reasonable because the exothermicity of Reactions \ref{react:h2s}--\ref{react:ph2} is much greater than the surface temperature.

For the surface reactions of \ce{PH3} with H atoms, the experiments were conducted on porous ASW only at 10 K \citep{nguyen21}.
Nonetheless, the $\pcd$ for the \ce{PH3} + H system can be constrained because we can constrain the binding energy distribution of atomic H by modeling the \ce{H2S} + H system.
Similar to the \ce{H2S} + H system, that the desorption probability for \ce{PH3} per incident H atom does not depend on the ice structure \citep{nguyen21}.

\subsection{Numerical setup} \label{sec:setup}
A square grid is considered with an $n \times n$ square lattice providing, at each point, an adsorption site for chemical species on porous ASW.
Each site thus has four neighboring sites.
Periodic boundary conditions are used.
Six different chemical species---atomic H, \ce{H2}, \ce{H2S}, HS, \ce{PH3}, and \ce{PH2}---are considered in our models.
These different chemical species are assumed to share adsorption sites, whereas the binding energy of each species is assumed to differ even at the same site.
Initially, some of the adsorption sites are occupied by \ce{H2S} or \ce{PH3}, as in the experiments by \citet{oba19} and \citet{nguyen21}.
The fluxes of H atoms onto the ASW surface are $5.7\times10^{13}$ cm$^{-2}$ s$^{-1}$ and $2.2\times10^{14}$ cm$^{-2}$ s$^{-1}$ for simulations of the \ce{H2S} + H experiments and the \ce{PH3} + H experiments, respectively.
These fluxes are consistent with those used in \citet{oba19} and \citet{nguyen21}.
We set the flux of \ce{H2} to be two-thirds of that of H atoms because the H atoms are produced from \ce{H2} with a dissociation probability of $>$60 \% \citep{oba18}. 
Refer to Section \ref{model:stick} for the sticking coefficients adopted in our models.

We consider thermal hopping and thermal desorption of H atoms and \ce{H2} on the ASW surface, whereas those of \ce{H2S}, HS, \ce{PH3}, and \ce{PH2} are ignored because their binding energies are relatively high ($\gtrsim$2000 K, see Section \ref{model:edes} for details).
If an H atom is adsorbed onto or hops to a site already occupied by another adsorbate, the two can react.
For surface reactions, \ce{H2} formation via recombination of two H atoms and Reactions \ref{react:h2s}--\ref{react:ph2} are considered.
Table \ref{table:react} summarizes the exothermicity, the activation energy barrier (if one exists), and the rate coefficient (s$^{-1}$) for Reactions \ref{react:h2s}--\ref{react:ph2}.
If a \ce{H2} molecule is adsorbed onto or hops to a site already occupied by S-bearing or P-bearing species, the two do not react and we assume that \ce{H2} is on the topmost layer at the site, with the adsorbate underneath.
In this case, we assume that the binding energy of the \ce{H2} is the same as that on porous ASW.
We do not allow the adsorbate underneath a \ce{H2} molecule to experience any chemical processes until the \ce{H2} molecule desorbs or hops to a neighboring site.
The presence of \ce{H2} on the surface thus reduces the efficiency of surface reactions in our models.
The slowing of surface reactions due to the presence of \ce{H2} in laboratory experiments has been reported by \citet{hama15}, who studied addition reactions of H atoms with solid benzene.
Multiple layers of solid \ce{H2} cannot be created even at 10 K \citep[e.g.,][]{hama12}.
Then we assume that neither \ce{H2} nor H atoms can be adsorbed onto or hop to a site already occupied by either \ce{H2} or atomic H in our simulations.
The exception is the adsorption of atomic H onto a site occupied by another atomic H, which instantly leads to the formation of \ce{H2}.

\begin{table*}
\caption{Summary of reaction properties}
\label{table:react}
\begin{center}
\begin{tabular}{ccccc}
\hline\hline        
 & (\ref{react:h2s}) $\react{H2S}{H}{HS}{H2}$ &  (\ref{react:hs}) $\reacti{HS}{H}{H2S}$ &  (\ref{react:ph3}) $\react{PH3}{H}{PH2}{H2}$ &  (\ref{react:ph2}) $\reacti{PH2}{H}{PH3}$ \\
\hline
Reaction energy (kJ mol$^{-1}$) & -58 \tablenotemark{{\rm a}}       & -374 \tablenotemark{{\rm a}}      &  -81.5 \tablenotemark{{\rm c}}    & -337.5 \tablenotemark{{\rm c}}                                            \\ 
Activation energy (K)    &              1560 \tablenotemark{{\rm b}}       & barrierless                                         &  1280 \tablenotemark{{\rm c}}                      & barrierless                                          \\
Rate coefficient  (s$^{-1}$) &         $8.2\times10^6$ \tablenotemark{{\rm d}}   & -                                         &  $8.7\times10^7$ \tablenotemark{{\rm e}}                        & -                                                  \\
\hline
\end{tabular}
\end{center}
\tablecomments{
$^{a}$ Gas-phase values \citep{oba18}. \\
$^{b}$ \citet{lamberts17}. \\
$^{c}$ Calculated values using a 20 water cluster model \citep{molpeceres21}. \\
$^{d}$ Value at 55 K, which is the lowest temperature studied in quantum chemistry calculations by \citet{lamberts17}. Note that quantum tunneling becomes important at $\lesssim$300 K and that the rate constants at 70 K and 55 K are almost identical \citep{lamberts17}. \\
$^{e}$ Value at 50 K, which is the lowest temperature studied in quantum chemistry calculations by \citet{molpeceres21}.
}
\end{table*}

\begin{figure}[t!]
\plotone{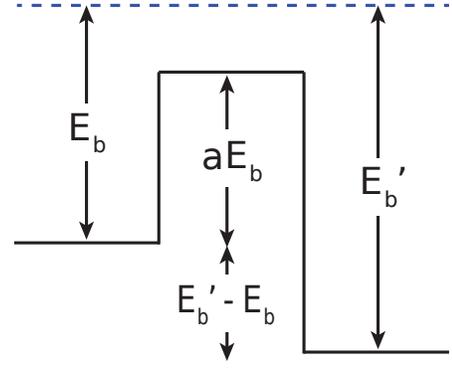}
\caption{Schematic of the hopping activation barrier from one site to another when the two sites have different binding energies. Based on Fig. 11 in \citet{cazaux17}.}
\label{fig:hopping}
\end{figure}

The thermal desorption rate (s$^{-1}$) of species $i$ depends on its binding energy on the surface ($\edes(i)$),
\begin{equation}
k_{\rm des}(i) = \nu_{\rm des} \exp(-\edes(i)/kT_s),
\end{equation}
where $\nu_{\rm des}$ is the characteristic attempt frequency for thermal desorption \citep[$1.8\times10^{12}$ s$^{-1}$;][]{amiaud15}, $T_s$ is the surface temperature, and $k$ is the Boltzmann constant.
The binding energies adopted in our models are described in Section \ref{model:edes}.
Similarly, the surface diffusion rate for species $i$ via thermal hopping depends on the hopping activation energy ($\ehop$):
\begin{equation}
k_{\rm diff,\,hop}(i) = \nu_{\rm hop} \exp(-\ehop/kT_s), \label{eq:khop}
\end{equation}
where $\nu_{\rm hop}$ is the attempt frequency for hopping in one direction, and we arbitrarily assume that $\nu_{\rm des} = \nu_{\rm hop}$.
The activation energy for hopping from a site with a binding energy of $\edes$ to another site with a binding energy of $\edes'$ is given as \citep[][see Fig. \ref{fig:hopping}]{cazaux17}:
\begin{equation}
\ehop (\edes \rightarrow \edes') = \alpha \times {\rm min}(\edes,\,\,\edes') + {\rm max}(0,\,\,\edes - \edes'),
\end{equation}
where $\alpha$, which is the hopping-to-binding energy ratio, is set to 0.65 for both atomic H and \ce{H2}, referring to the recommendation based on calculations of the diffusivity of atomic H on ASW \citep{Asgeirsson17}.
Given the expression for $\ehop$, the thermal hopping rate exhibits microscopic reversibility, i.e., $k_{\rm diff,\,hop}(\edes \rightarrow \edes')/k_{\rm diff,\,hop}(\edes' \rightarrow \edes) = \exp[-(\edes-\edes')/kT_s]$ \citep{cuppen13}.
Although Eq. \ref{eq:khop} is often used in kMC simulations \citep[e.g.,][]{cuppen07,garrod13}, a different expression is proposed for the rate of hopping between adsorption sites with different potential-energy depths \citep{mate20}.
We tested some models adopting Eq. 14 in \citet{mate20} instead of Eq. \ref{eq:khop} and confirmed that the simulation results are not sensitive to the expression of the hopping rate.
In our fiducial models, we do not consider the surface diffusion of atomic H by quantum tunneling. 
This effect will be discussed in Section \ref{seq:qt}.

\epsscale{0.8}
\epsscale{1.0}

For surface reactions, both the Langmuir--Hinshelwood mechanism and the Eley--Rideal mechanism are considered.
In our model, when two species arrive at the same adsorption site and can undergo a chemical reaction, we do not follow the details of the competition but determine immediately which process---reaction, hopping, or desorption---occurs by comparing their rates \citep[e.g.,][]{cuppen07,chang12}.
For barrierless reactions, we assume that the reaction occurs before desorption and hopping, with a probability of unity. 
For reactions with an activation barrier, the probability of the reaction occurring before one of the species (here, atomic H) leaves to the site, where the two species meet, is given by
\begin{equation}
p_{\rm A+H} = \frac{k_{\rm A+H}}{k_{\rm A+H} + k_{\rm des}({\rm H}) + \Sigma_{i=1}^{4}k^i_{\rm diff,\,hop}({\rm H})}, \label{eq:competition}
\end{equation}
where A is either \ce{H2S} or \ce{PH3} and $k_{\rm A+H}$ is the reaction rate coefficient.
The term $\Sigma_{i=1}^{4}k^i_{\rm diff,\,hop}$ originates from our assumption of a square lattice arrangement of surface sites; each site has four neighboring sites to which atomic H in the site can hop.
Eq. \ref{eq:competition} is implemented in the Monte Carlo algorithm as follows \citep{chang07}: 
A random number between 0 and 1 is generated. 
If this number is smaller than $p_{\rm A+H}$, the reaction occurs. 
Otherwise, thermal desorption or hopping of the atomic H occurs on the basis of their rates and the random number.

\begin{figure}[t!]
\plotone{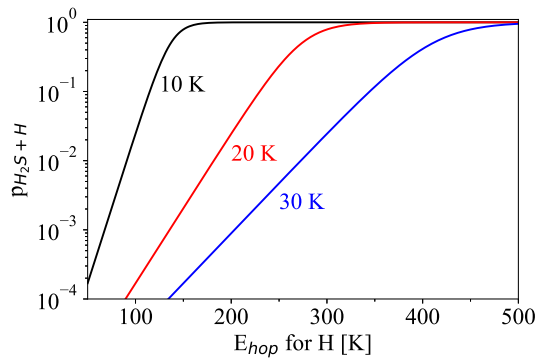}
\caption{Probability for reactions to occur before thermal hopping of atomic H to a neighboring adsorption site occurs (Eq. \ref{eq:competition}), expressed as a function of the hopping activation energy at a surface temperature of 10 K (black), 20 K (red), and 30 K (blue). 
In this figure, $\ehop$ is assumed to be the same for the four neighboring sites.
Thermal desorption of atomic H is neglected because its timescale is generally longer than the timescale for thermal hopping.
}
\label{fig:preact}
\end{figure}

Figure \ref{fig:preact} shows $p_{\rm H_2S+H}$ as functions of $\ehop$ for atomic H at 10 K, 20 K, and 30 K under the assumption that $\ehop$ is the same for hopping to the four neighboring sites.
The value for parameter $k_{\rm H_2S+H}$ was taken from \citet{lamberts17}, who studied Reaction \ref{react:h2s} using quantum chemical calculations (Table \ref{table:react}).
At 10 K, $p_{\rm H_2S+H}$ is unity (i.e., the reaction is effectively barrierless) when $\ehop({\rm H}) \gtrsim 150$ K.
This threshold $\ehop$ is higher for higher $T_s$ because $k_{\rm diff,\,hop}$ for a given $\ehop$ increases with increasing $T_s$: it is 300 K and 450 K at $T_s = 20$ K and 30 K, respectively.
Notably, as a result of the loop of \ce{H2S}--HS interconversion via Reactions \ref{react:h2s} and \ref{react:hs}, the \ce{HS}/\ce{H2S} abundance ratio approaches $p_{\rm H_2S+H}$.

After the surface reactions, the products are released from the surface via chemical desorption with a probability of $\pcd$ (see Section \ref{model:dp}).
A random number between 0 and 1 is generated; if this number is smaller than the $\pcd$, the products are instantly released from the surface.
Throughout this paper, we use the term ``chemical desorption rate" as the rate of a surface reaction multiplied by the $\pcd$.

\subsection{Chemical parameters} \label{model:chem}
\subsubsection{Sticking coefficient} \label{model:stick}
The sticking coefficient depends on various parameters, such as the incident kinetic energy of colliding atoms/molecules, surface temperature, and the surface composition.
The temperature of atomic H (and \ce{H2}) deposited on the ASW was 100 K in the experiments \citep{oba19,nguyen21}.
In our models, the sticking coefficient for atomic H is taken from \citet{dupuy16}, who determined the sticking coefficient for atomic H on ASW for various incident kinetic energies and surface temperatures on the basis of classical molecular dynamics (MD) simulations.
In our models, the sticking coefficient is set to 0.65 at $T_s = 10$ K, and 0.6 at 20 K and 30 K.
These sticking coefficients are used in both cases where atomic H lands on empty sites (i.e., \ce{H2O}) and cases where H lands on either \ce{H2S} or \ce{PH3}.
The sticking coefficient for \ce{H2} is assumed to be the same as that for atomic H.

\subsubsection{Binding energy} \label{model:edes}
\begin{figure}[t!]
\plotone{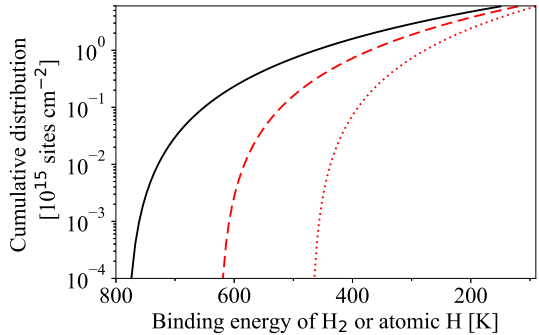}
\caption{Cumulative distribution of the H$_2$ binding energy adopted in the present work (black line).
Note that the minimum value of $\edes$ for \ce{H2} is treated as a free parameter.
The red dashed line and the red dotted lines represent the cumulative distribution of the binding energy for atomic H in models with $f=0.8$ and $f=0.6$, respectively.
When $f=1.0$, the binding energy distribution for atomic H is identical to that for \ce{H2}. 
See the text for details.
}
\label{fig:edes_dist}
\end{figure}

\citet{amiaud06, amiaud15} constrained the binding energy distribution of \ce{H2} on porous ASW on the basis of temperature-programmed desorption (TPD) experiments.
They found that the binding energy distribution can be described by a polynomial function, adopting $\nu_{\rm des}$ = $1.8\times10^{12}$ s$^{-1}$ \citep{amiaud15}:
\begin{align}
   g_{\rm H_2}(\edes) = \begin{cases} \label{eq:eb_dist}
    a(E_{\rm b}^{\rm max} - \edes)^{1.6} & (\edesmin \leq \edes < E_{\rm b}^{\rm max})  \\
   0 & (otherwise) 
  \end{cases}  
\end{align} 
where $a$ = $4.7\times10^{11}$ sites cm$^{-2}$ meV$^{-2.6}$ and $E_{\rm b}^{\rm max} =$ 67.5 meV ($\sim$780 K) \citep{amiaud15}.
Notably, the functional form of $g_{\rm H_2}$ indicates that the number of adsorption sites with smaller $\edes$ is greater.
The lower bound of $\edes$ ($\edesmin$) was not constrained well by the TPD experiments because \ce{H2} was deposited on ASW at a surface temperature of 10 K; even if adsorption sites with very low $\edes$ exist, they would not be probed by the TPD experiments because of efficient thermal desorption even at 10 K.
For example, the timescale of thermal desorption from sites with $\edes = 250$ K is only $\sim$0.1 s at 10 K.
The minimum value of $\edes$ constrained by their experiments was $\sim$30 meV ($\sim$350 K) \citep{amiaud06}; however, we extrapolate $g_{\rm H_2}(\edes)$ to lower $\edes$.
In the present work, $\edesmin$ is treated as a free parameter and we test three values: $\edesmin$ = 150 K, 250 K, and 350 K.
According to the numerical investigation of \citet{molpeceres20}, who used MD simulations, approximately 75\% of the adsorption sites for \ce{H2} on ASW have an $\edes$ in the range between 150 K and 450 K and sites with even lower $\edes$ exist, although the peak binding energy distribution is $\sim$300 K in their simulations.
Because the intermolecular potential energy of a \ce{H2}--\ce{H2O} dimer is $\sim$120 K \citep{zhang91}, the model with $\edesmin$ = 150 K corresponds to the situation where the majority of sites have an $\edes$ as low as the intermolecular energy of a dimer. 
We also tested some models in which the binding energy distributions for \ce{H2} (and atomic H) follow a Gaussian distribution, but found no clear advantage over using a polynomial distribution (see Appendix for details). 
The cumulative distribution for the H$_2$ binding energy is shown in Fig. \ref{fig:edes_dist}.

The atomic H binding energy obtained from experiments is not well constrained because of the high reactivity of atomic H.
In this work, we assume that the binding energy of atomic H is given by $f\edes({\rm H_2})$, where $f$ is a scaling factor.
\citet{wakelam17} have reported that the binding energy of stable molecules on ASW, as determined from TPD experiments, is proportional to the intermolecular potential energy between a dimer of the molecule and \ce{H2O} (see their Fig. 1).
The intermolecular energy between a dimer of atomic H and \ce{H2O} is $\sim$70 K, whereas that between a dimer of \ce{H2} and \ce{H2O} is $\sim$120 K \citep{zhang91} (i.e., the former is $\sim$0.6 times smaller than the latter).
We varied $f$ in the range $0.6 \leq f \leq 1.0$ in the present work.
The cumulative distribution of the atomic H binding energy is shown by red lines in Fig. \ref{fig:edes_dist}.
Notably, the binding energy distribution for atomic H in our models is controlled by two independent parameters: $\edesmin({\rm H_2})$ and $f$.
The former parameter controls the extent of $\edes({\rm H})$, whereas the latter parameter controls the maximum value of $\edes({\rm H})$, which is given by $f E_{\rm b}^{\rm max}({\rm H_2})$.
Notably, we assume that the presence of \ce{H2S} or \ce{PH3} does not affect the binding energies of atomic H and \ce{H2} on porous ASW.

At the beginning of each kMC simulation, $\edes({\rm H_2})$ at each lattice site is assigned randomly, but it follows $g_{\rm H_2}$.
The $\edes({\rm H})$ at each lattice site is set to $f\edes({\rm H_2})$; that is, \ce{H2} and atomic H share adsorption sites and shallow (deep) potential sites for \ce{H2} are also shallow (deep) for atomic H.

According to TPD experiments, the binding energy of \ce{H2S} on water ice is 2700 K \citep{collings04,garrod06,jimenez-escobar11}.
To the best of our knowledge, no experimental measurements of the binding energy of HS, \ce{PH3}, or \ce{PH2} on water ice have been reported.
However, theoretically calculated binding energies for HS, \ce{PH3}, or \ce{PH2} on water ice are available: 2700 K for HS \citep{wakelam17}, 2200 K for \ce{PH3}, and 1800 K for \ce{PH2} \citep[mean values;][]{molpeceres21}.
\citet{sil21} and \citet{nguyen21} also calculated the binding energy of \ce{PH3} on water ice and reported a value similar to that of \citet{molpeceres21}.
In our model, we neglect thermal desorption and thermal hopping of \ce{H2S}, HS, \ce{PH3}, and \ce{PH2}.
Given the relatively high binding energy of these molecules, this assumption should be valid under the experimental conditions ($T_s \leq 30$ K).

The integration of $g_{\rm H_2}(\edes)$ from $E_{\rm b}^{\rm max}$ to $\edesmin$ gives the site density ($N_{\rm site}$) on porous ASW. 
The $N_{\rm site}$ on porous ASW is known to be greater than that on flat crystalline water ice (10$^{15}$ molecules per cm$^2$) \citep[][]{kimmel01,hidaka08}.
$N_{\rm site}$ is $\sim$6$\times10^{15}$, $\sim$4$\times10^{15}$, and $\sim$2$\times10^{15}$ sites cm$^{-2}$ when $\edesmin$ = 150 K, 250 K, and 350 K, respectively.
We use different number of grids to unify the resolution of our models; 
$n$ is determined to satisfy the relation, $N_{\rm site}/(n\times n) \approx 2\times10^{11}$ sites cm$^{-2}$.
In the models with $\edesmin$ = 150 K, 250 K, and 350 K, we use $161\times161$, $128\times128$, and $98\times98$ square lattices, respectively.

In the experiments performed by \citet{oba19}, $7\times10^{14}$ molecules cm$^{-2}$ of \ce{H2S} were deposited on ASW.
In this case, the initial surface coverage of \ce{H2S} ($7\times10^{14}$/$N_{\rm site}$) is $\sim$12\%, $\sim$18\%, and $\sim$35\% in our models with $\edesmin$ = 150 K, 250 K, and 350 K, respectively.
The initial positions of \ce{H2S} molecules on the grid are randomly determined at the beginning of each simulation.

\subsubsection{Chemical desorption probability} \label{model:dp}
The chemical desorption probabilities per reaction event for Reaction \ref{react:h2s} (denoted as $\pcds$) and for Reaction \ref{react:ph3} ($\pcdp$) are treated as free parameters in the present work.
In the experiments by \citet{oba18,oba19}, it was not possible to distinguish between the chemical desorption of \ce{H2S} upon Reaction \ref{react:h2s} and the chemical desorption of \ce{HS} upon Reaction \ref{react:hs}.
If the ratio of the binding energy of a reaction product to the energy released by the reaction is a key parameter
for determining $\pcd$ \citep[e.g.,][]{garrod07,minissale16}, then the $\pcd$ for the hydrogenation reactions would be greater than those for the hydrogen abstraction reactions (see Table \ref{table:react}). 
In the present work, the $\pcd$ of HS upon Reaction \ref{react:hs} ($\pcd({\rm HS})$) is always set to be lower than the $\pcds$ by a factor of 10.
Notably, our model results do not depend on the ratio of $\pcd({\rm HS})$ to $\pcds$ but do depend on their sum because the chemical desorption of \ce{H2S} and HS occurs via the loop of Reactions \ref{react:h2s} and \ref{react:hs}.
Similarly, the $\pcd$ of \ce{PH2} for Reaction \ref{react:ph2} ($\pcd({\rm PH_2})$) is set to be lower than that for Reaction \ref{react:ph3} ($\pcdp$) by a factor of 10.
Throughout this work, $\pcd$ is assumed to be independent of both $T_s$ and the adsorption sites.

For simplicity, we assume that $\pcd$ of \ce{H2}, which can be formed by the recombination of H atoms and the hydrogen abstraction reactions, is unity.
This assumption does not affect our results because \ce{H2} can be directly adsorbed on the surface and because the flux of \ce{H2} is as high as that of H atoms; the maximum formation rate of \ce{H2} on the surface (i.e., one-half of the adsorption rate of atomic H) is comparable to the adsorption rate for \ce{H2}.
Thus, the main source of \ce{H2} on the surface is direct adsorption.

\begin{table}
\caption{Free parameters for the kinetic Monte Carlo simulations}
\label{table:kmc}
\begin{center}
\begin{tabular}{ccc}
\hline\hline
Parameter              & Values \\
\hline
$\edesmin({\rm H_2})$\tablenotemark{{\rm a}}       & 150 K (161$^2$) - 250 K (128$^2$) - 350 K (98$^2$)  \\
$f$                  &  0.6 - 0.7 - 0.8 - 0.85 - 1.0  \\
$\pcd$             & 1 \% - 2 \% - 3 \% - 5 \% - 10 \% - 20 \% \\
\hline
\end{tabular}
\end{center}
\tablecomments{
$^{a}$ The values in parentheses are the number of grids used in models with different $\edesmin({\rm H_2})$ value. 
}
\end{table}

Our model has the three free parameters (i.e., $\edesmin({\rm H_2})$, $f$, and $\pcd$); they are summarized in Table \ref{table:kmc}.
In total, we ran 90 different models in which the three free parameters were varied.
For each model, we ran the simulations five times with different random seeds and then took an average to reduce fluctuations in the calculation results.
We then found the best ``by-eye" fit model among the models.
This model was considered to be the best fit to the experiments.

\section{Results} \label{sec:result}
\subsection{H$_2$S + H} \label{sec:h2s_result}
\subsubsection{Fiducial model} \label{sec:specific}
\begin{figure*}[ht!]
\plotone{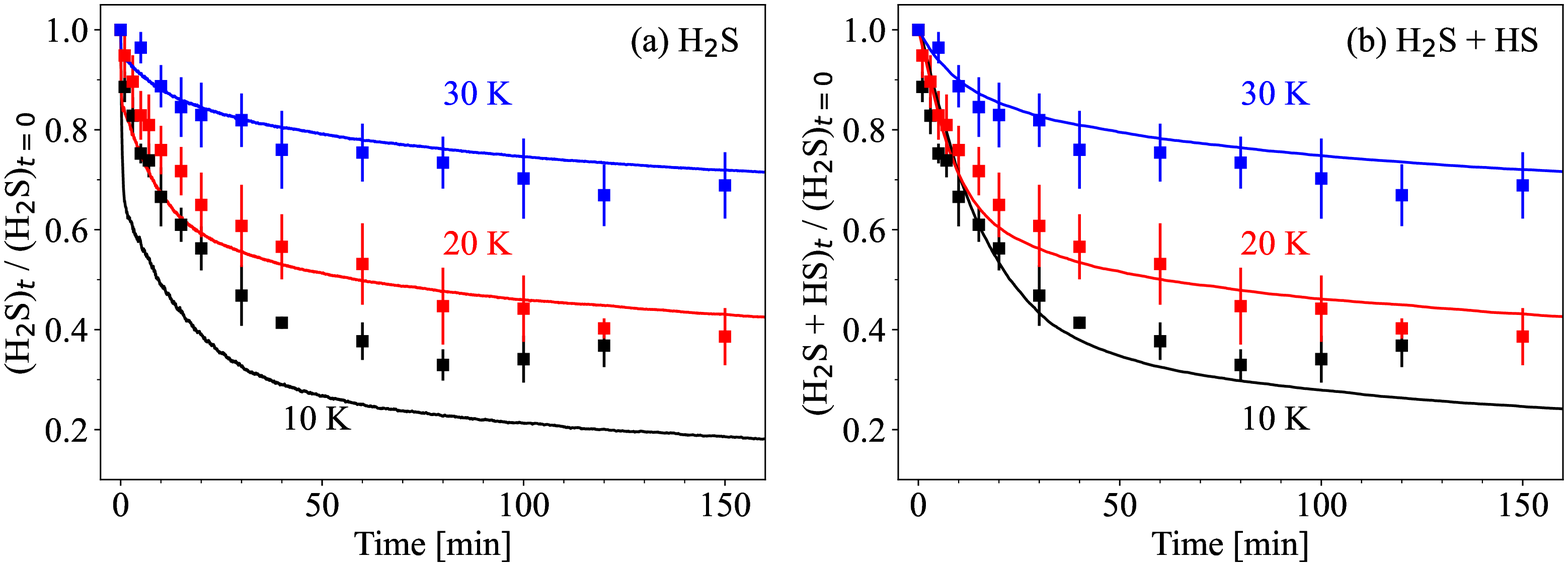}
\caption{\ce{H2S} abundance (left panel) and total abundance of \ce{H2S} and HS (right panel) on ASW, normalized by the initial \ce{H2S} abundance and plotted as functions of the H-atom exposure time in the model with $\edesmin({\rm H_2})=150$ K, $f = 0.85$, and $\pcds = 3$\% (lines).
Square symbols represent experimental results reported by \citet{oba19} and based on FTIR spectroscopy.
Black, red, and blue represent surface temperatures of 10 K, 20 K, and 30 K, respectively.
}
\label{fig:best_h2s}
\end{figure*}

\begin{figure}[t!]
\plotone{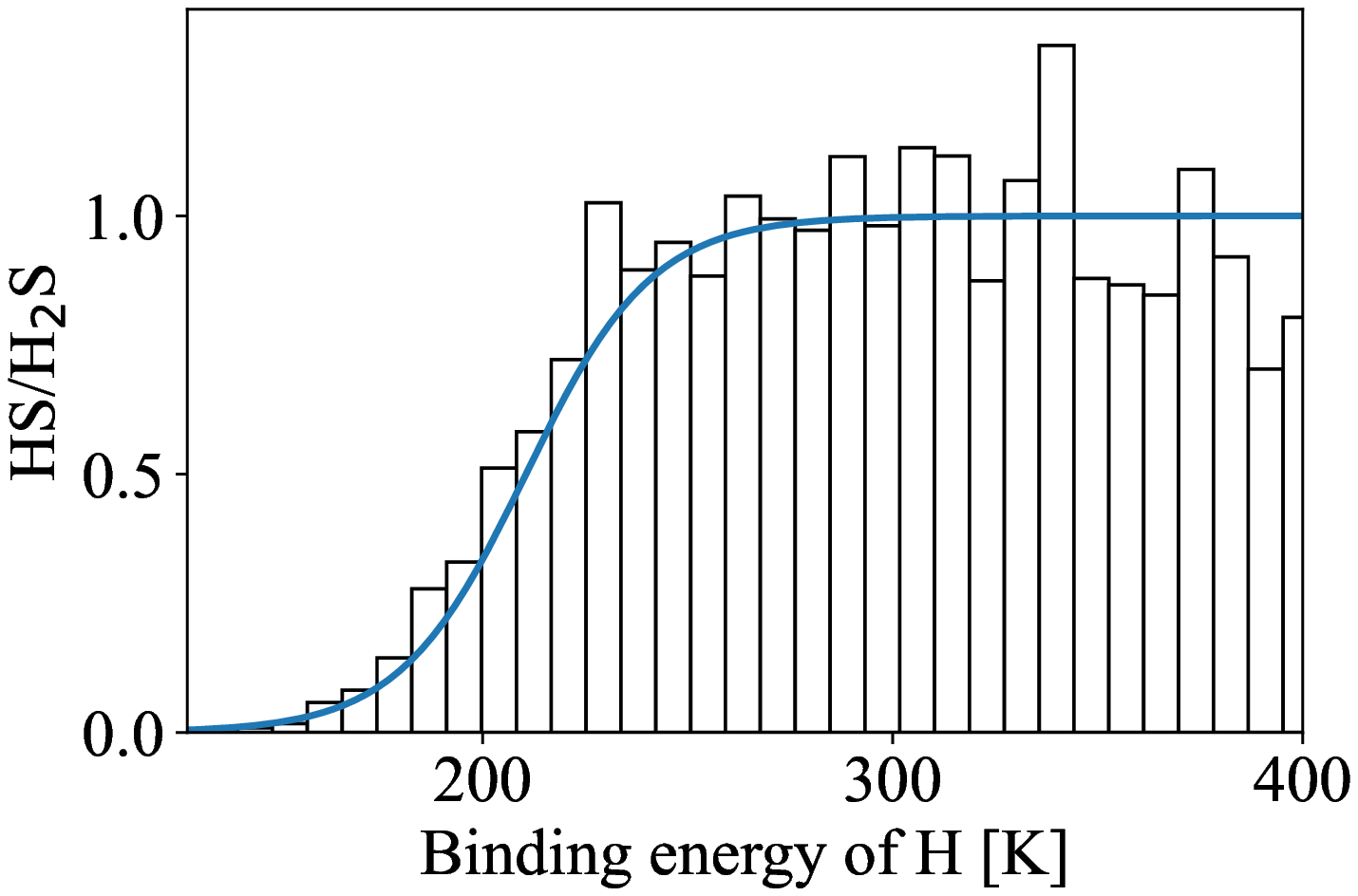}
\caption{
HS/\ce{H2S} abundance ratio in adsorption sites with different $\edes({\rm H})$ in the model with $\edesmin({\rm H_2})=150$ K, $f = 0.85$, and $\pcds = 3$\% at $t=1$ min (black bars).
For comparison, $p_{\rm H_2S+H}$ is shown as a blue line.
In the evaluation of $p_{\rm H_2S+H}$, $\ehop({\rm H})$ is assumed to be 0.65$\edes({\rm H})$.
}
\label{fig:hs_h2s_ratio}
\end{figure}

We first consider the results from one specific model to understand how the binding energy distribution affects the surface chemistry.
The left panel of Figure \ref{fig:best_h2s} shows the \ce{H2S} abundance on ASW normalized by the initial \ce{H2S} abundance as a function of the H-atom exposure time in the model with $\edesmin({\rm H_2})=150$ K, $f = 0.85$, and $\pcds = 3$ \% (hereafter referred to as the fiducial model).
The results show that $\sim$30 \% of \ce{H2S} is lost within the first 1 min in the model at 10 K, which is clearly inconsistent with the experiments.
Although the decrease in the \ce{H2S} abundance at later times ($\gtrsim$1 min) is due to chemical desorption, this rapid loss of \ce{H2S} is not due to chemical desorption; as a result of the loop of \ce{H2S}--HS interconversion via Reactions \ref{react:h2s} and \ref{react:hs}, the abundance of \ce{H2S} decreases, whereas the abundance of HS increases.
At $T_s = 10$ K, the probability that Reaction \ref{react:h2s} occurs per encounter of atomic H with \ce{H2S} is unity (i.e., $p_{\rm H_2S+H} = 1$ and the reaction is effectively barrierless) when $\ehop({\rm H}) \gtrsim 150$ K (see Fig. \ref{fig:preact}).
Because both Reactions \ref{react:h2s} and \ref{react:hs} are (effectively) barrierless in sites with $\ehop({\rm H}) \gtrsim 150$ K, the abundances of HS and \ce{H2S} become similar as a result of the \ce{H2S}--HS interconversion in these sites (Fig. \ref{fig:hs_h2s_ratio}).
This \ce{H2S}--HS interconversion should occur before substantial amounts of \ce{H2S} and HS are released from the surface by chemical desorption as long as $\pcds \ll 1$.
Indeed, this rapid destruction and formation of \ce{H2S} and HS, respectively, via \ce{H2S}--HS interconversion are observed in all our models at 10 K, irrespective of the values of $\edesmin({\rm H_2})$, $f$, and $\pcds$.
The degree of \ce{H2S} loss depends on the binding energy distribution of atomic H (i.e., the fraction of sites with $\ehop({\rm H}) \gtrsim 150$ K); however, more than 20\% of the initial amount of \ce{H2S} is converted to HS within 1 min in all of our models at 10 K.

We conclude that the rapid destruction and formation of \ce{H2S} and HS, respectively, via \ce{H2S}--HS interconversion is inevitable at 10 K.
In principle, if all adsorption sites have $\ehop({\rm H}) \ll 150$ K ($\edes({\rm H}) \ll 230$ K as $\ehop({\rm H}) \sim 0.65\edes({\rm H})$), then the abundance of HS would be negligible because $p_{\rm H_2S+H} \ll 1$ for all sites, and the HS/\ce{H2S} abundance ratio would therefore be much lower than unity at 10 K (see Figure \ref{fig:hs_h2s_ratio}).
In such models, however, reproducing the results of the experiments at 20 K and 30 K is difficult because efficient thermal desorption (e.g., the resident time of atomic H on the surface would be $<$10$^{-7}$ s at 20 K and even shorter at 30 K) prevents the occurrence of Reaction \ref{react:h2s} and the chemical desorption of \ce{H2S} and HS would be negligible in the experimental timescale.
The rapid destruction and formation of \ce{H2S} and HS, respectively, observed in the model at 10 K becomes much less obvious in the models at 20 K and 30 K (left panel of Fig. \ref{fig:best_h2s}) because the thermal hopping rate of atomic H increases exponentially
with increasing $T_s$.
Thus, $p_{\rm H_2S+H}$ is much less than unity for the majority of the adsorption sites; i.e., the HS/\ce{H2S} abundance ratio $\approx p_{\rm H_2S+H} \ll 1$.

\citet{oba19} confirmed the chemical desorption of \ce{H2S} (and HS) by both FTIR spectroscopy in atomic H deposition experiments and QMS measurements in TPD experiments after H-atom deposition.
During H-atom deposition, a decrease was observed in the magnitude of the band at 2570 cm$^{-1}$, which was assigned to an S--H stretching band of \ce{H2S}.
However, distinguishing between \ce{H2S} and \ce{HS} by FTIR is difficult because of overlap of the absorption bands \citep{jimenez-escobar11}.
In the TPD experiments after H-atom deposition, a decrease in the \ce{H2S} amount compared with the initial \ce{H2S} amount deposited on the ASW surface was found by QMS.
The QMS measurements confirmed that no S-bearing species except \ce{H2S} were desorbed; however, both \ce{H2S+} and \ce{HS+} were produced as fragment ions of \ce{H2S}, which could mask the existence of HS.
Thus, we cannot rule out the possibility that some HS along with \ce{H2S} existed on the ASW in the experiments.

In the present work, as a working hypothesis, we interpret the 2570 cm$^{-1}$ band as the sum of the S--H stretching modes of \ce{H2S} and HS rather than solely as the stretching mode of \ce{H2S}.
We also assume that the band intensity for \ce{H2S} and HS on ASW are the same because no literature value has been reported for HS.
Under these assumptions, in the rest of this paper, we focus on the total abundance of S (i.e., \ce{H2S} + HS) on the ASW surface rather than the \ce{H2S} abundance in our simulations.
In the right panel of Fig. \ref{fig:best_h2s}, the vertical axis represents the total abundance of \ce{H2S} and \ce{HS} normalized by the initial abundance of \ce{H2S} at a given time.
In this case, we do not observe a rapid decrease at early times ($\lesssim30$ s) at 10 K, whereas the results at 20 K and 30 K are similar between the two panels.

\begin{figure*}[ht!]
\plotone{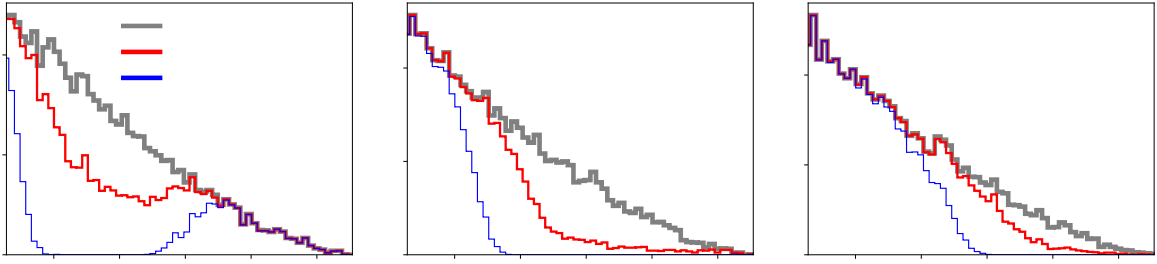}
\caption{Number of adsorption sites occupied by either \ce{H2S} or HS at $t = 0$ min (gray), at 10 min (red), and at 150 min (blue) 
in the models with $f = 0.85$, $\edesmin=150$ K, and $\pcds = 3$\% as functions of the binding energy of atomic H in the adsorption sites.
The left, middle, and right panels show the model for a surface temperature of 10 K, 20 K, and 30 K, respectively.
Note that the majority of adsorption sites are empty and that these empty sites are not included in the figure.
}
\label{fig:h2s_dist}
\end{figure*}

In the experiments, the abundance of \ce{H2S} (and HS) on ASW decreased with time and eventually became almost constant in the experimental timescale (Fig. \ref{fig:best_h2s}). 
This nonlinear behavior reflects the binding energy distributions of atomic H and \ce{H2} on ASW.
Figure \ref{fig:h2s_dist} shows the number of adsorption sites occupied by either \ce{H2S} or HS at $t = 0$ min (gray), 10 min (red), and 150 min (blue) in the fiducial model at $T_s$ = 10 K (left), 20 K (middle), and 30 K (right) as functions of $\edes({\rm H})$.
Because the surface coverage of \ce{H2S} is $\sim$12 \% at $t = 0$ min (Section \ref{model:edes}), the majority of adsorption sites are empty and these empty sites are not included in the figure.
At 10 K and at $t = 150$ min, \ce{H2S} and HS are populating adsorption sites with binding energies $\edes({\rm H})$ following a bimodal distribution ($\edes({\rm H}) \lesssim 150$ K and $\edes({\rm H}) \gtrsim 400$ K).
In the deep sites ($\edes({\rm H}) \gtrsim 450$ K, which corresponds to $\edes({\rm H_2}) \gtrsim 530$ K as $\edes({\rm H}) =  0.85\edes({\rm H_2})$ in the fiducial model), \ce{H2S} and HS are buried by \ce{H2};
because \ce{H2} does not easily hop to neighboring sites, further reactions of \ce{H2S} and HS with atomic H are hindered, as described in Section \ref{sec:setup}.
The hopping timescale for \ce{H2} from a site with $\edes({\rm H_2}) = 530$ K to a site with $\edes({\rm H_2}) \lesssim 450$ K exceeds the duration of the experiments (150 min).
In the shallow sites ($\edes({\rm H}) \lesssim 150$ K), $p_{\rm H_2S+H}$ is less than unity, which slows the loop of \ce{H2S}--HS interconversion;
the rate of the chemical desorption is thereby reduced.
\ce{H2S} only in the sites with $\edes({\rm H}) \lesssim 250$ K and $\edes({\rm H}) \lesssim 350$ K remains on the surface after 150 min at 20 K and 30 K, respectively.
\ce{H2} readily hops from the deepest sites to shallower sites and is thermally desorbed at 20 K and 30 K; essentially no \ce{H2} remains on ASW.
A bimodal distribution is therefore not observed at 20 K and 30 K.

We can roughly estimate the threshold value of $\edes({\rm H})$ ($\edesthresh({\rm H})$); most \ce{H2S} in sites with a binding energy lower than $\edesthresh({\rm H})$ remains on ASW after 150 min of H-atom deposition, avoiding chemical desorption.
The flux of H atoms is $5.7\times10^{13}$ cm$^{-2}$ s$^{-1}$ with a sticking probability of $\sim$0.6; thus, the fluence (i.e., the time integral of the flux) of H atoms in 150 min is $\sim3\times10^{17}$ molecules cm$^{-2}$.
Because $7\times10^{14}$ molecules cm$^{-2}$ of \ce{H2S} are initially present on ASW, the average number of H atoms available to desorb one \ce{H2S} molecule via chemical desorption is $\sim$400.
Let us assume that adsorbed H atoms are consumed either by the loop of Reactions \ref{react:h2s}--\ref{react:hs} in sites with $\edes$ or by thermal desorption.
Under this assumption, the probability that Reaction \ref{react:h2s} occurs before thermal desorption of a H atom is given by $k_{\rm H_2S+H}/(k_{\rm H_2S+H} + k_{\rm des}({\rm H}))$.
Then, by solving $400[k_{\rm H_2S+H}/(k_{\rm H_2S+H} + k_{\rm des}({\rm H}))] \pcds = 1$, we obtain $\edesthresh({\rm H})$ to be $\sim$100 K, 200 K, 300 K at $T_s = 10$ K, 20 K, and 30 K, respectively, when $\pcds = 3$\%.
These values are consistent with the numerical results corresponding to $T_s = 20$ K and 30 K (see Fig. \ref{fig:h2s_dist}).
In the numerical simulation at 10 K, \ce{H2S} and HS in adsorption sites even with $\edes({\rm H}) > \edesthresh({\rm H})$ remain on the surface.
The lack of agreement between the previously discussed simple argument and the numerical results at 10 K indicates the importance of trapping H atoms in deep-potential sites, followed by \ce{H2} formation via recombination with another H atom.
This effect can reduce the number of H atoms available for the \ce{H2S}--HS interconversion loop.
Even for $\edes({\rm H})$ = 660 K (i.e., the deepest site), the timescale of hopping to a neighboring site is around 0.001 s at 20 K, 
which is comparable to the average adsorption timescale of one atomic H in our model, and the hopping timescale is even shorter at 30 K.
Then the trapping of atomic H in deep sites is only important at 10 K.

Taken together, the nonlinear behavior of the total abundance of \ce{H2S} and HS on ASW can be understood as follows:
At early times ($\lesssim$10 min), chemical desorption occurs at adsorption sites, where the loop of \ce{H2S}--HS interconversion can occur most efficiently (i.e., $p_{\rm H_2S+H} \sim 1$).
At later times, 
the rate of chemical desorption is reduced,
because the remaining \ce{H2S} is in sites less favorable for the \ce{H2S}--HS interconversion loop; that is, either $p_{\rm H_2S+H} < 1$ or \ce{H2S} is buried by \ce{H2}, hindering further reactions.
Eventually, \ce{H2S} only in sites with $\edes({\rm H})$ lower than $\edesthresh({\rm H})$ remains on the ASW.
At 10 K, \ce{H2S} (and HS) also remain in deep-potential sites, where the presence of \ce{H2} hinders surface reactions.

The previous discussion reveals that the binding energy distribution for atomic H can be directly constrained on the basis of the 20 K and 30 K experiments because the fraction of \ce{H2S} remaining on the ASW corresponds to the fraction of adsorption sites with $\edes({\rm H})$ lower than $\edesthresh({\rm H})$.
As previously mentioned, $\edesthresh({\rm H})$ is $\sim$200 K and 300 K at  $T_s = 20$ K and 30 K, respectively.
$\edesthresh({\rm H})$ depends logarithmically on $\pcds$; i.e., the dependence of $\edesthresh({\rm H})$ on $\pcds$ is weak.
The abundance of \ce{H2S} becomes $\sim$40\% and $\sim$70\% of the initial \ce{H2S} abundance at 150 min in the 20 K and 30 K experiments, respectively. 
Thus, $\sim$70\% of the adsorption sites on porous ASW should have $\edes({\rm H}) \lesssim 300$ K, whereas $\sim$40\% of the adsorption sites should have $\edes({\rm H}) \lesssim 200$ K.
\citet{hama12} found in their experiments that adsorption sites for atomic H on the ASW surface can be categorized into three groups: shallow ($\ehop({\rm H}) < 18$ meV $\sim$ 210 K), middle ($\ehop({\rm H}) = 22$ meV $\sim$ 260 K), and deep ($\ehop({\rm H}) > 30$ meV $\sim$ 350 K) potential sites.
Although the fraction of each of the three sites was not constrained well in their work, the shallow sites were found to be dominant on the surface.
If we assume a hopping-to-binding energy ratio of 0.65 \citep{Asgeirsson17}, the binding energy distribution constrained as previously described indicates that $\sim$70\% of the adsorption sites have $\ehop({\rm H}) \lesssim 0.65\times300 \sim 200$ K, consistent with the diffusion barriers constrained by \citet{hama12}.

\subsubsection{Lower limit of H$_2$S desorption probability} \label{sec:ll}
As previously discussed, the initial decrease of the \ce{H2S} and HS abundances at $\lesssim10$ min in the experiments is likely due to chemical desorption from adsorption sites, where $p_{\rm H_2S+H} = 1$ and, thus, the loop of \ce{H2S}--HS interconversion can occur most efficiently.
We expect that reproducing the experimental results at $\lesssim$10 min using models would be easier than reproducing the experimental results at $>$10 min, where chemical desorption occurs even in sites with $p_{\rm H_2S+H} < 1$.
Thus, as a first step, attempting to constrain the $\pcds$ while focusing only on the experimental results at $\lesssim$10 min would be worthwhile.
For this purpose, we additionally ran models with a single value of $\ehop({\rm H})$ for all adsorption sites.
We varied $\ehop({\rm H})$ in the range where $p_{\rm H_2S+H} = 1$ (i.e., $\ehop({\rm H}) > 150$ K at $T_s = 10$ K).
In these models, we neglected thermal desorption of atomic H for simplicity, whereas we considered \ce{H2} formation by recombination of two H atoms.
The other parameters were the same as those described in Section \ref{model:chem}.
We here focus on the 10 K experiments because thermal desorption of atomic H should occur efficiently at 20 K and 30 K, invalidating the aforementioned assumption.

\begin{figure}[t!]
\plotone{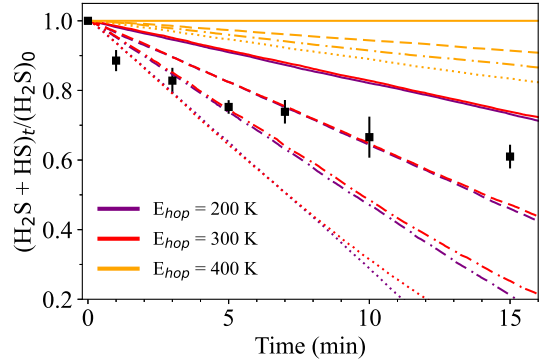}
\caption{Temporal evolution of the \ce{H2S} abundance in the model with a single value of $\ehop({\rm H})$ at 10 K.
Solid, dashed, dash-dotted, and dotted lines represent the models with $\pcds$ of 1\%, 2\%, 3\%, and 4\%, respectively.
The models with different $\ehop({\rm H})$ are represented in different colors.
In all the models, thermal desorption of atomic H is neglected. 
Black squares show the experimental data \citep{oba19}.
}
\label{fig:h2s_low}
\end{figure}

Figure \ref{fig:h2s_low} shows the total abundance of \ce{H2S} and HS normalized by the initial \ce{H2S} abundance as functions of the H-atom exposure time at 10 K, where $\ehop({\rm H})$ and $\pcds$ were varied.
When $\ehop({\rm H}) = 400$ K or larger, $\ehop({\rm H})$ and $\pcds$ degenerate (i.e., we cannot constrain $\pcds$).
However, when $\ehop({\rm H}) = 200$ K and 300 K, the results do not depend on $\ehop({\rm H})$ but do depend on $\pcds$ because the diffusion of H atoms by thermal hopping is fast and almost all the adsorbed H atoms are consumed by either Reaction \ref{react:h2s} or Reaction \ref{react:hs} before another H atom is adsorbed onto the ASW surface.
That is, Reactions \ref{react:h2s} and \ref{react:hs} are adsorption-limited and the number of \ce{H2S} and HS released from the surface is simply given by $\pcds$ multiplied by the number of adsorbed H atoms.
In the experiments, some adsorbed H atoms might have been trapped in deep-potential sites and/or desorbed thermally; these effects are not included in the models with a single value of $\ehop$ but indeed occur in the models where the binding energy distribution is taken into account.
Then, what we can constrain here is the lower limit of $\pcds$ by comparing the models with $\ehop \leq 300$ K with the experiments at $t \lesssim 10$ min.
If we consider the experimental results only at $t \leq 5$ min, the lower limit of $\pcds$ is $\sim$3\%.
If we consider the experimental results at $t \leq 10$ min, the lower limit of $\pcds$ is $\sim$2\%.
As a conservative choice, we consider the lower limit of $\pcds$ as 2\%.

\subsubsection{Constraining the H$_2$S desorption probability}
\begin{figure}[t!]
\plotone{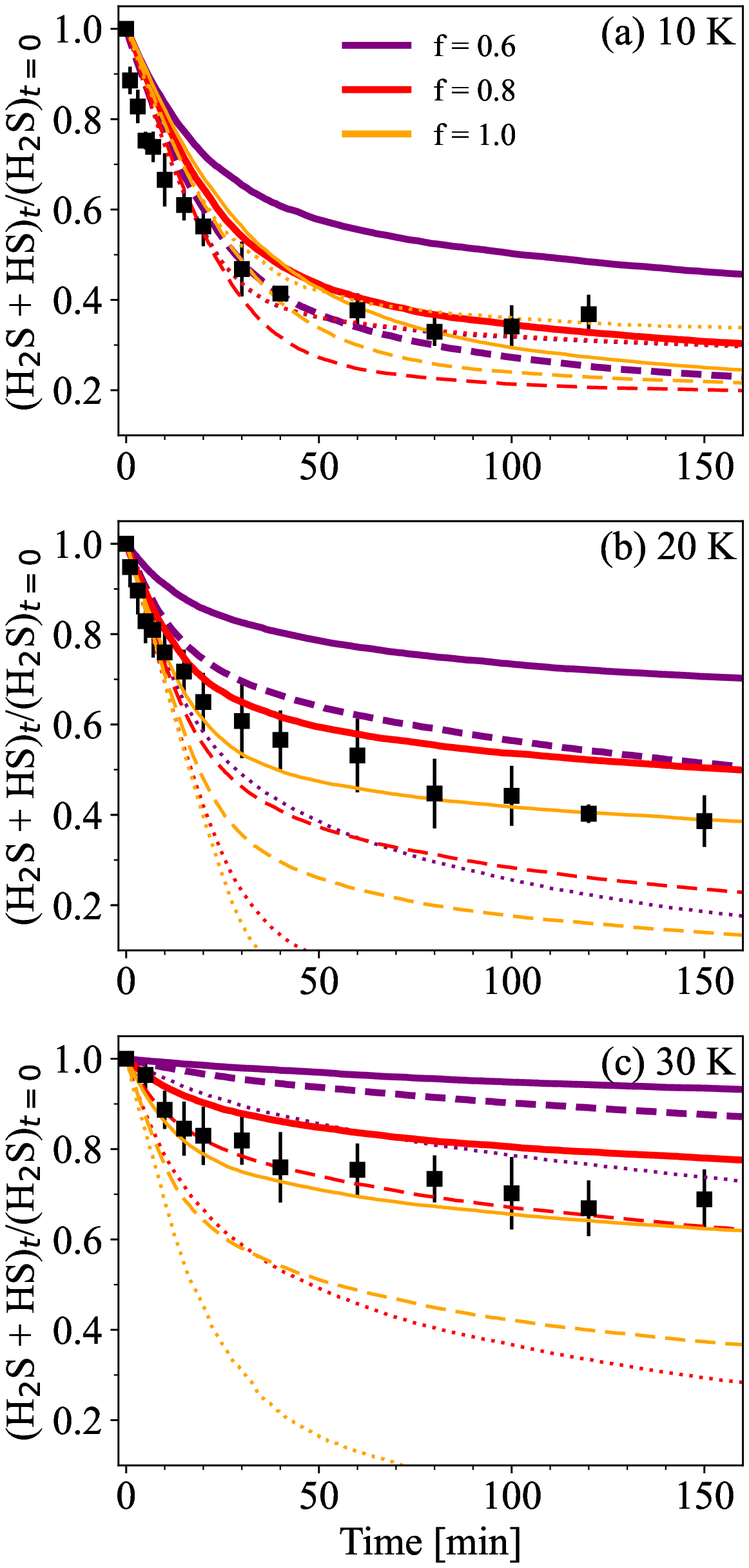}
\caption{Total S (\ce{H2S} + HS) abundance with respect to the initial \ce{H2S} abundance on porous ASW at 10 K (panel a), 20 K (panel b), and 30 K (panel c) as functions of the H-atom exposure time in the kMC simulations (lines) and in the experiments by \citet{oba19} (black squares).
Solid, dashed, and dotted lines represent the models with $\edesmin({\rm H_2})$ values of 150 K, 250 K, and 350 K, respectively.
The models with different $\ehop({\rm H})$ values are represented in different colors.
In all the simulations, $\pcds$ is set to 2\%.
}
\label{fig:h2s}
\end{figure}

Figure \ref{fig:h2s} shows the total amount of \ce{H2S} and \ce{HS} divided by the initial amount of \ce{H2S} at 10 K, 20 K, and 30 K as functions of the H-atom exposure time in the models with $\pcds = 2$\%, where the other free parameters, $f$ and $\edesmin({\rm H_2})$, were varied.
The model results are sensitive to $f$ and $\edesmin({\rm H_2})$ at 20 K and 30 K, whereas the impact of the two parameters is less significant at 10 K.
At 20 K and 30 K, \ce{H2S} and \ce{HS} in adsorption sites with higher $\edesmin({\rm H})$ are preferentially lost from the surface (Fig. \ref{fig:h2s_dist}).
Lower values of $f$ and $\edesmin({\rm H_2})$ lead to a decrease in the fraction of deeper sites.
As a result, the chemical desorption rate is reduced with decreasing $f$ and/or $\edesmin({\rm H_2})$, as evident in Figure \ref{fig:h2s}.
At 10 K, \ce{H2S} and \ce{HS} become populating adsorption sites with binding energies $\edesmin({\rm H})$ following a bimodal distribution (Fig. \ref{fig:h2s_dist}).
Lower values of $f$ and $\edesmin({\rm H_2})$ lead to a decrease in the fraction of deep sites (i.e., the chemical desorption rate tends to increase), whereas they lead to an increase in the fraction of shallow sites (i.e., the chemical desorption rate tends to decrease).
As a result, the 10 K model is less sensitive to $f$ and $\edesmin({\rm H_2})$ than the 20 K and 30 K models.

Because 2\% is the lower limit of $\pcds$, the models that overestimate the amount of \ce{H2S} and \ce{HS} desorbed from the ASW surface compared to the experiments are ruled out.
Examining the 20 K and 30 K models, we can rule out the majority (6 of 9) of the models shown here.
The three models not yet ruled out are those with $\edesmin({\rm H_2}) = 250$ K and $f = 0.6$, with $\edesmin({\rm H_2}) = 150$ K and $f = 0.6$, and with $\edesmin({\rm H_2}) = 150$ K and $f = 0.8$ (shown by thick lines in Fig. \ref{fig:h2s}).
Notably, the three models share the common feature of a low atomic H binding energy; more than one-half of the adsorption sites have a low binding energy for atomic H ($\lesssim$300 K, see Fig. \ref{fig:edes_dist}) and, thus, a low hopping barrier ($\lesssim$200 K).

Among the three models, we disfavor the model with $\edesmin({\rm H_2}) = 250$ K and $f = 0.6$ because, although the model moderately well reproduces the experimental results at $T_s = 10$ K, too much \ce{H2S} and HS remain on the surface at 30 K compared to the experiments; the experiments at 10 K, 20 K, and 30 K are difficult to fit simultaneously by varying $\pcds$.
We thus have two remaining possibilities: (i) the model with $f=0.6$ and $\edesmin({\rm H_2}) = 150$ K is more favorable and $\pcds$ is much higher than the lower limit value of 2\% because too much \ce{H2S} and HS remains on the surface compared with the experiments at all the investigated temperatures, or (ii) the model with $f=0.8$ and $\edesmin({\rm H_2}) = 150$ K is more favorable and $\pcds$ is close to and slightly greater than 2\% because the model somehow underestimates chemical desorption at all the investigated temperatures.

To explore the first possibility, we tested additional models, varying $f$ from 0.6 to 0.8 and varying $\pcds$ from 5\% to 20\% but fixing $\edesmin({\rm H_2}) = 150$ K.
The results for these models are shown in Fig. \ref{fig:grid2} in the appendix.
We find that reproducing the 10--30 K experiments simultaneously using these models with a relatively high $\pcds$ of $>$5\% is difficult; the models at 10 K overestimate the amount of \ce{H2S} and HS desorbed from the surface compared with the experiments, or the models at 30 K underestimate the amount of \ce{H2S} and HS desorbed from the surface compared with the experiments.

Therefore, we conclude that the second possibility is more favorable; i.e., the $\pcds$ is close to the lower limit value of 2\%.
After some exploration, we found that the model with $f=0.85$, $\edesmin({\rm H_2}) = 150$ K, and $\pcds = 3$\% (i.e., the fiducial model) reasonably well reproduces the 10 K, 20 K, and 30 K experimental results simultaneously (see the right panel of Fig. \ref{fig:best_h2s}).
We confirmed that the reduced $\chi^2$ for this model (= 1.85) is the minimum among all the models applied in the present work.
We consider this model as the best-fit model.

As noted in Section \ref{model:dp}, our model results depend on the sum of $\pcds$ and $\pcd({\rm HS})$, but do not depend on the ratio of $\pcd({\rm HS})$ to $\pcds$, which is assumed to be 0.1 throughout the present work.
Thus, strictly speaking, the constraint applied here is that the sum of $\pcds$ and $\pcd({\rm HS})$ is 3.3 \%.
Nevertheless, $\pcds$ can be reasonably expressed as $\sim$3\% because $\pcds$ would be greater than $\pcd({\rm HS})$ according to the theoretical models for chemical desorption (see Section \ref{sec:theory}).
Finally, we also ran models in which the binding energy distributions of atomic H and \ce{H2} are given by a Gaussian distribution rather than by a polynomial distribution; we found no clear advantage to using a Gaussian distribution (see the appendix).

Our best-fit model underestimates the total abundance of \ce{H2S} and HS remaining on the ASW surface at 10 K compared with the experiments, although it better reproduces the 20 K and 30 K experimental results.
This underestimation might indicate that the presence of \ce{H2} on the surface lowers the $\pcd$.
The ASW surface is partly covered by \ce{H2} at 10 K, whereas essentially no \ce{H2} remains on the surface at 20 K and 30 K.
\citet{minissale14} found in their experiments that the chemical desorption probability of \ce{O2} upon the reaction between two O atoms decreases with increasing coverage of \ce{O2} on oxidized graphite.
The presence of an adsorbed species might enhance the dissipation of the excess energy produced by chemical reactions, resulting in a lowering of the $\pcd$ \citet{minissale14}.
Such an effect of an adsorbed species (\ce{H2} in our case) on the $\pcd$ is not considered in our models.

\subsection{Constraining the PH$_3$ desorption probability}
\begin{figure}[t!]
\plotone{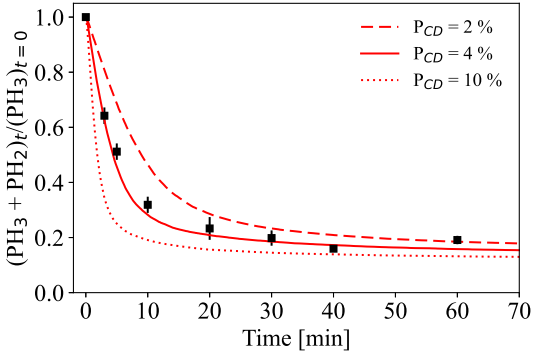}
\caption{Relative abundances of phosphorous (\ce{PH3} + \ce{PH2}) on porous ASW at 10 K as functions of the H-atom exposure time in the kMC simulations (lines) and in the experiments by \citet{nguyen21} (black squares).
Solid, dashed, and dotted lines represent the models with $\pcdp$ of 4\%, 2\%, and 10 \%, respectively.
In all the simulations, $\edesmin({\rm H_2})$ and $f$ are set to 150 K and 0.85, respectively.
}
\label{fig:ph3}
\end{figure}

We here constrain the $\pcd$ for \ce{PH3} upon Reaction \ref{react:ph2}.
As previously described, the experiments of \ce{H2S} + H at 10 K, 20 K, and 30 K are reasonably well reproduced by the model with $f = 0.85$ and $\edesmin({\rm H_2}) = 150$ K.
Therefore, in this subsection, we fixed the parameters $f$ and $\edesmin$ to these values.
As in Section \ref{sec:h2s_result}, we discuss the total amount of \ce{PH3} and \ce{PH2} on the surface divided by the initial amount of \ce{PH3}, again because the rapid conversion of \ce{PH3} to \ce{PH2} is inevitable at $T_s = 10$ K in our models; the rate of Reaction \ref{react:ph3} is even greater than that of Reaction \ref{react:h2s} (Table \ref{table:react}).
Figure \ref{fig:ph3} compares the experimental results with the model, where the $\pcdp$ was varied.
Notably, the H atom flux was approximately four times greater in the experiments involving \ce{PH3} + H \citep{nguyen21} than in those involving \ce{H2S} + H \citep{oba19}.
We find that the model with $\pcdp = 4$\% reasonably well reproduced the experiments at 10 K.

\section{Discussion} \label{sec:discuss}
\subsection{Uncertainties in the flux of atomic H} \label{discuss:stick}
Thus far, the flux of H atoms was set to $5.7\times10^{13}$ cm$^{-2}$ s$^{-1}$ and $2.2\times10^{14}$ cm$^{-2}$ s$^{-1}$ in the \ce{H2S} + H model and in the \ce{PH3} + H model, respectively. 
These fluxes were estimated by \citet{oba19} and \citet{nguyen21}.
The authors did not discuss the uncertainties in the estimation of the atomic H flux in their experiments; however, empirically, the absolute uncertainties can be as large as 50\% (private communication).
Here, we explore how the flux uncertainty affects the constraints on $\pcd$.

We tested two additional models for the \ce{H2S} + H system, varying the H atom flux.
In one model, the H atom flux was set to $3.8\times10^{13}$ cm$^{-2}$ s$^{-1}$ (50\% lower than the fiducial value); in the other model, it was  $8.6\times10^{13}$ cm$^{-2}$ s$^{-1}$ (50\% higher).
The other parameters were the same as in our fiducial model: $\edesmin({\rm H_2}) = 150$ K, $f=0.85$, and $\pcds = 3$\%.
The model with the 50\% lower H flux and $\pcds = 3$\%, and the model with the fiducial H flux and $\pcds = 1.5$\% gave almost identical results, whereas the model with the 50\% higher H flux and $\pcds = 3$\% and the model with the fiducial H flux and $\pcds = 4.5$\% gave almost identical results.
Thus, the uncertainty in the H atom flux is inversely linearly transferred to the uncertainty in the $\pcds$, likely because the system is adsorption-limited rather than diffusion-limited.
Considering the empirical uncertainty in the H atom flux (50\%), we conclude that the $\pcds$ is $3 \pm 1.5$\%, whereas the $\pcdp$ is $4 \pm 2$\%.

\subsection{Diffusion of atomic H by quantum tunneling} \label{seq:qt}
In our fiducial models, thermal hopping is treated solely as a mechanism of surface diffusion.
\citet{Asgeirsson17} theoretically studied the diffusion of atomic H on amorphous ice, considering both thermal hopping and quantum tunneling, and showed that tunneling is important only for $ T_s \lesssim 10$ K.
\citet{kuwahata15} experimentally showed that tunneling dominates thermal hopping for atomic H on crystalline water ice at 10 K.
However, tunneling is less important for the diffusivity of atomic H on ASW, where greater heterogeneity (i.e., a greater range in $\edes$(H)) exists compared to the heterogeneity of crystalline water ice surfaces.
The rationale is that, even though H atoms can tunnel between shallow sites, the rate-limiting step for long-range diffusion is the transition out of deep sites, which still requires thermal activation \citep{smoluchowski83,kuwahata15,Asgeirsson17}.
These experimental and theoretical studies indicate that tunneling has limited importance for long-range diffusion of atomic H on ASW at $\geq$10 K.
On the other hand, tunneling diffusion might play a role in determining the reaction probability because it is determined by the competition between reaction and short-range diffusion (Eq. \ref{eq:competition}).

\begin{figure*}[t!]
\plotone{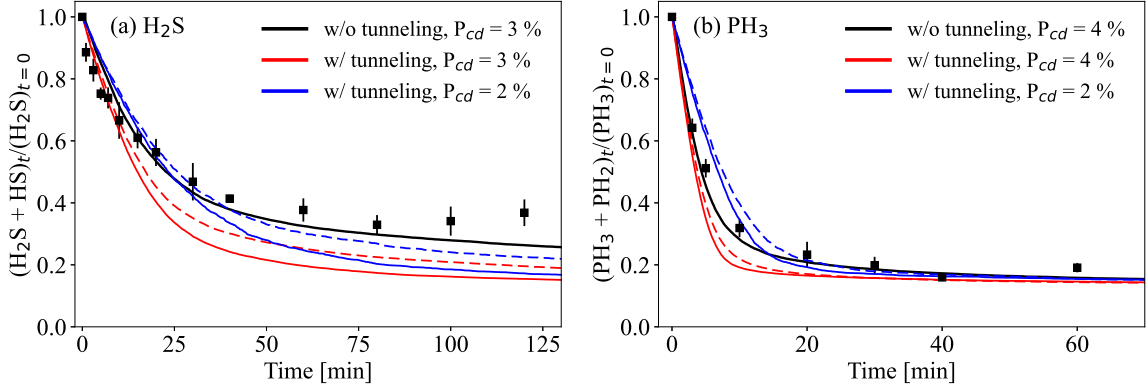}
\caption{
Total abundance of \ce{H2S} and HS (left panel) and total abundance of \ce{PH3} and \ce{PH2} as functions of the H atom exposure time in models with tunneling diffusion with a barrier thickness of 1 \AA  (red and blue solid lines) and in the models with tunneling diffusion with a barrier thickness of 2 \AA (red and blue dashed lines) at 10 K.
In all the models, $\edesmin({\rm H_2})=150$ K and $f = 0.85$.
Our best-fit models constrained as described in Section \ref{sec:result} are shown by black solid lines for comparison.
Square symbols represent the experimental results reported by \citet{oba19} and \citet{nguyen21}.
}
\label{fig:tunnel}
\end{figure*}

To check the effect of tunneling diffusion on our results, we tested additional models in which the tunneling diffusion of atomic H was considered.
\citet{lamberts14} proposed tunneling rates for exothermic reactions based on arguments of microscopic reversibility.
Because the diffusion rates should also obey microscopic reversibility, referring to \citet{lamberts14}, we calculate the tunneling diffusion rates for atomic H between two sites with different binding energies ($\edes < \edes'$) as
\begin{align}
&k_{\rm diff,\,qt}(\edes \rightarrow \edes') = \nu \exp \left(-\frac{2d}{\hbar}\sqrt{2m_{\rm H}\alpha\edes}\right), \\
&k_{\rm diff,\,qt}(\edes' \rightarrow \edes) \nonumber \\
  & \,\,\,= k_{\rm diff,\,qt}(\edes \rightarrow \edes') \exp\left(-\frac{\edes'-\edes}{kT_s}\right), \label{eq:qm2}
\end{align}
where surface sites are assumed to be separated by a rectangular barrier of thickness $d$ and $m_{\rm H}$ is the mass of atomic H.
Because the barrier thickness is unknown, we here consider two values: $d = 1$ \AA{} and 2 \AA. 
Notably, when atomic H moves from a deeper site to a shallower site, the part of activation energy barrier that corresponds to the difference in the binding energies of the two sites should be overcome thermally (see Fig. \ref{fig:hopping}).
This consideration is accounted for by the factor $\exp(-(\edes'-\edes)/kT_s)$ in Eq. \ref{eq:qm2}.
We compared the tunneling rate and thermal hopping rate and chose the higher of the two as the diffusion rate.

Figure \ref{fig:tunnel} shows the effect of tunneling diffusion at $T_s = 10$ K when $d=1$ \AA{} (red solid line) and $d=2$ \AA{} (red dashed line).
Compared with our best-fit model constrained as described in Section \ref{sec:result} (black solid line), tunneling increases the fraction of \ce{H2S} and HS released to the gas phase by chemical desorption to some extent.
As a result, the model with tunneling diffusion underestimates the amount of \ce{H2S} and HS remaining on the ASW surface after 25 min compared with the experimentally observed amount.
Even if we adopt $\pcds = 2$\%, which is the lower limit value, the model with tunneling underestimates the amount of \ce{H2S} and HS remaining on the ASW surface (blue lines).
As discussed in Section \ref{sec:specific}, some \ce{H2S} and HS in sites with $\edes({\rm H}) > \edesthresh({\rm H}) \sim 100$ K remain on the surface at 10 K in our fiducial model (see Fig. \ref{fig:h2s_dist}) because of trapping of H atoms in deep-potential sites, followed by the formation of \ce{H2} via recombination with another H atom.
Because of the tunneling, the effect of trapping is reduced and almost all of the \ce{H2S} and HS in sites with $\edes({\rm H}) \sim 150$ K are released to the gas phase by chemical desorption in the model with tunneling diffusion.
A similar effect is observed in the \ce{PH3} + H model at 10 K.
At 20 K and 30 K, the results obtained using the models with and without tunneling diffusion are almost identical (not shown).

When tunneling diffusion is considered, 
the model with $\pcds = 2$ \% better reproduces the \ce{H2S} + H experimental results at $T_s = 10$ K, 20 K, and 30 K rather than the model with $\pcds = 3$ \%, 
assuming  $\edesmin({\rm H_2}) = 150$ K and $f = 0.85$.
Note, however, that the reduced $\chi^2$ for this model is larger than that for our best-fit model constrained in Section \ref{sec:result} ($\sim$4 versus $\sim$2).
For the \ce{PH3} + H system, the model with $\pcds = 3$ \% better reproduces experimental results rather than the model with $\pcdp = 4$ \%, when tunneling diffusion is considered.
The reduced $\chi^2$ for this model is similar to that for our best-fit model constrained in Section \ref{sec:result}. 
Therefore, slightly lower $\pcd$ than that constrained in Section 3 is favorable, when tunneling diffusion is considered.

\subsection{Comparison with theoretical models of chemical desorption} \label{sec:theory}
Chemical desorption is caused by the energy released by reactions; after an exothermic surface reaction, some of the excess energy is dissipated into products' translational energy in the direction perpendicular to the surface, leading to desorption \citep[e.g.,][]{fredon17}. 
The efficiency of chemical desorption depends on (i) the excess energy released by reaction ($E_{\rm react}$), (ii) the binding energy of products to the surface, (iii) the fraction of excess energy that remains in the products and is not lost to the solid surface, and (iv) how the energy of the products is distributed among all degrees of freedom.
The last two parameters are currently uncertain, which limits our quantitative understanding of chemical desorption.
Several authors have suggested general formalisms of chemical desorption to be used in astrochemical models \citep{garrod07,minissale16,fredon21}.
These formalisms include free parameters that should be determined by laboratory experiments or/and quantum chemistry calculations.
Here, we compare the $\pcd$ constrained in the present work with that predicted by the theoretical models of chemical desorption in the literature.

\citet{garrod07} proposed the following formalism in the case of one-product reactions, applying Rice--Ramsperger--Kessel--Marcus theory:
\begin{align}
&P_{\rm cd,\,G07} = \frac{a\gamma}{1+a\gamma}, \\
&\gamma = \left(1-\frac{\edes}{E_{\rm react}} \right)^{s-1}, \label{eq:garrod}
\end{align}
where $\gamma$ is the probability for an energy $E > \edes$ to be present in the admolecule--surface bond and $s$ is the number of vibrational degrees of freedom, including binding to the surface.
The parameter $a = \nu/\nu_s$ is the ratio of the desorption attempt frequency to the frequency at which the reaction energy is lost to the surface.
The value of $a$ is unknown, and $a=0.01$ (i.e., $P_{\rm cd,\,G07} \sim 1$\%) is often assumed in astrochemical models \citep[e.g.,][]{garrod07,taquet14,furuya15}.
In the case of two-product reactions, chemical desorption is assumed to not occur \citep{garrod07}.

\citet{minissale16} proposed a different formulation from that of \citet{garrod07} to explain the results of their chemical desorption experiments.
They treated the energy dissipation as an elastic collision, where some of the kinetic energy is transferred to the surface from the product species, and assumed that excess energy that remains in the product species is spread equally over all degrees of freedom of the product species.
Their formulation with an effective surface mass of 130 amu reproduced the efficiency of chemical desorption for the studied reactions on graphite surfaces (i.e., rigid surfaces) within the margin of error.
However, the use of the formulation by \citet{minissale16} is questionable for ASW surfaces, where intermolecular interactions dominate and the collisions between admolecules and the surface are inelastic \citep{fredon17}.
Because our work is focused on chemical desorption on ASW, hereafter we do not discuss the formalism of \citet{minissale16}.

More recently, \citet{fredon21} proposed the following formulation based on classical MD simulations, where excess energy is transferred to a molecule adsorbed on ASW and the fate of the molecule is tracked \citep[see also][]{fredon17}:
\begin{align}
P_{\rm cd,\,F21} = \alpha \left(1- \exp \left(-\frac{\chi E_{\rm react} - \edes}{3\edes} \right)\right), \label{eq:fredon}
\end{align}
where $\alpha$ is an empirical factor of 0.5 and the factor of 3 in the denominator arises from the assumption that only translational excitation in the direction perpendicular to the surface can lead to desorption.
The free parameter $\chi$ describes the fraction of excess energy transferred to the translational excitation of products.
They proposed that $\chi$ depends on the number of product species; for product species $i$, $\chi$ is given by
\begin{align}
   \chi_i = \begin{cases} \label{eq:chi}
    \chi_1 & ({\rm one\,\,product\,\,reaction}),  \\
   \chi_2 \frac{m_j}{m_i + m_j} & ({\rm two\,\,product\,\,reaction}),
  \end{cases}  
\end{align}
where $m_i$ is the mass of product species $i$, and $m_j$ is the mass of another species produced by the reaction.
\citet{fredon21} conducted gas--ice astrochemical simulations using the rate equation method in conjunction with their formalism and found that their model reasonably well reproduces the observations of gas-phase (complex) organic molecules in dark clouds when $\chi_1 = 0.1$.
They also found that the gas-phase abundances of organic molecules are not sensitive to the choice of $\chi_2$ because surface reactions that are efficient at low temperatures ($\sim$10 K) involve atomic H or \ce{H2} and, thus, $\chi_i$ is small for organic molecules, which is more massive than atomic H and \ce{H2}.

\begin{figure*}[ht!]
\plotone{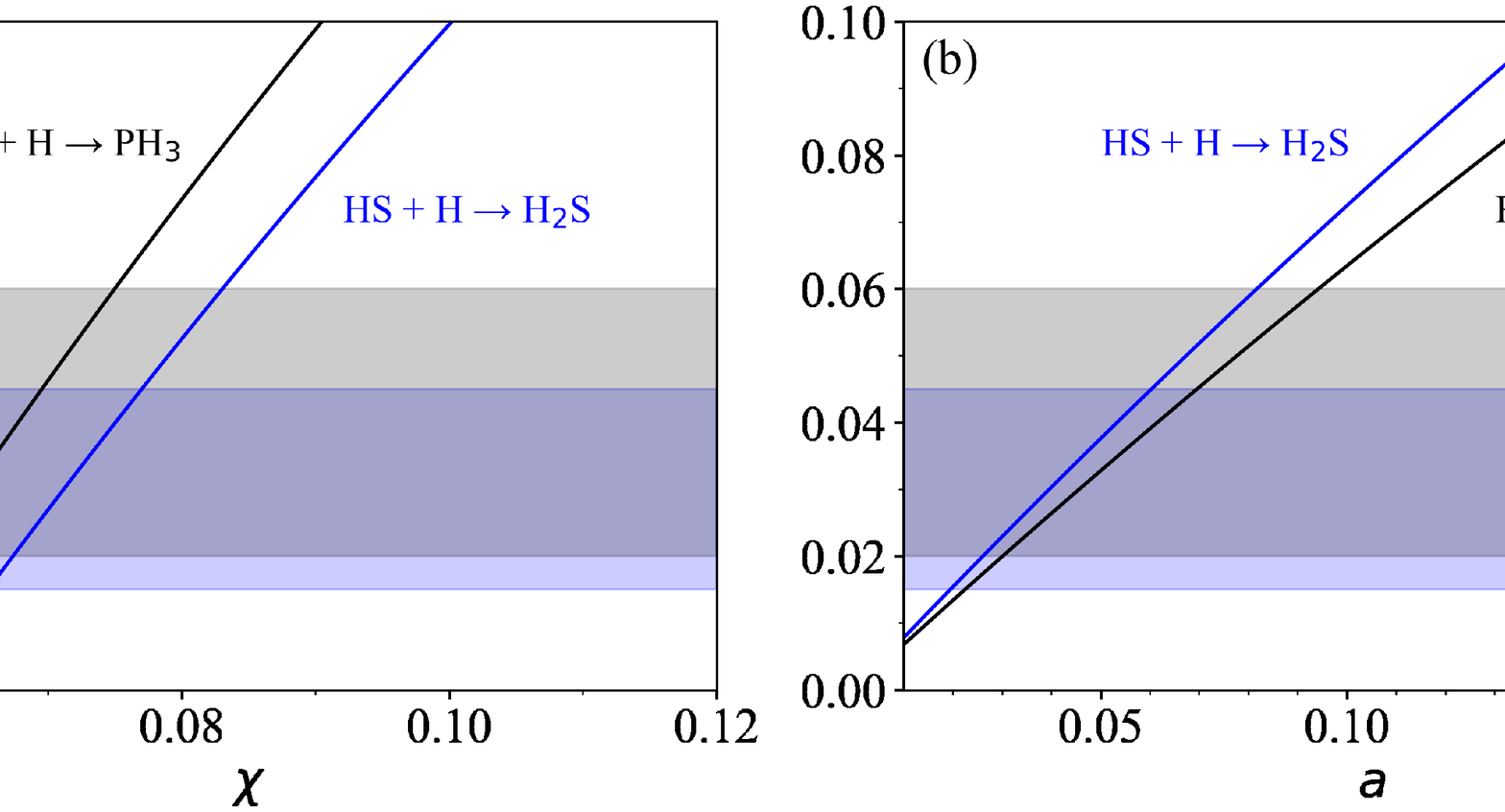}
\caption{
Chemical desorption probability of \ce{H2S} via Reaction \ref{react:hs} (blue line) and \ce{PH3} via Reaction \ref{react:ph2} (black line) predicted by the formulation of \citet[][]{fredon21} (Eq. \ref{eq:fredon}, panel a) and that by \citet{garrod07} (Eq. \ref{eq:garrod}, panel b).
$\pcds$ and $\pcdp$ constrained in the present work are shown by purple and gray areas, respectively.
}
\label{fig:analytic}
\end{figure*}

Figure \ref{fig:analytic} compares $\pcd$ for Reactions \ref{react:hs} and \ref{react:ph2} constrained as described in the present work with those predicted from the formulations proposed by \citet{fredon21} (left panel) and by \citet{garrod07} (right panel).
The binding energy of \ce{H2S} and \ce{PH3} was set to 2700 K and 2200 K, respectively \citep{collings04,molpeceres21}.
The energy released by the reactions is listed in Table \ref{table:react}.
The Fredon formulation predicts that $\pcds$ should be lower than $\pcdp$, whereas the Garrod formulation predicts the opposite trend.
Our results indicate that $\pcds$ is slightly lower than $\pcdp$; thus, the Fredon formulation is more consistent with our findings.
From a quantitative comparison, we find that the Fredon formulation with $\chi_1 \sim 0.07$ reproduces both $\pcds$ and $\pcdp$ constrained as described in the present work; i.e., approximately 7\% of the excess energy is transferred to the translational excitation of the products.
The Fredon formulation predicts that, even if $\chi_2$ is unity, the values of $\pcd({\rm HS})$ and $\pcd({\rm PH_2})$ are zero, adopting the \ce{HS} and \ce{PH2} binding energies at 2700 K and 1800 K, respectively.
These results support our assumption that $\pcd$ for the H addition reactions is higher than that for the H abstraction reactions.
If we adopt the Garrod formulation to explain $\pcds$ and $\pcdp$, the parameter $a$ should be $\sim$0.05, which is larger than the typically assumed value of 0.01.

Whether the Fredon formulation with $\chi_1 \sim 0.07$ provides reasonable estimates of $\pcd$ for other reaction systems is unclear.
\citet{pantaleone20} used ab initio MD simulations to study the fate of energy released by the reaction H + CO $\rightarrow$ HCO on crystalline water ice and found that 90\% of the reaction energy is instantly injected toward the water ice.
The fraction of excess energy transferred to the translational excitation should be less than 10\% in this case.
However, \citet{pantaleone21} studied the \ce{H2} formation on water ice surfaces using ab initio MD simulations and found that as much as two-thirds of the reaction energy is injected toward the water ice and the remaining energy is retained in the produced \ce{H2}.
Experimental and numerical studies of additional systems are required to draw any solid conclusions.

\subsection{Astrochemical implications}
Gaseous \ce{H2S} has been detected in various evolutionary stages of star and planet formation, cold dense clouds \citep[e.g.,][] {minh89,ohishi92,navarro-almaida20}, envelope around protostars \citep[e.g.,][]{blake94,wakelam04}, and protoplanetary disks \citep{phuong18,rivieremarichalar21}.
The formation of \ce{H2S} in the gas phase is inefficient because of the presence of endothermic reactions in the sequence of reaction pathways to convert \ce{S+} or atomic S into \ce{H2S} \citep[e.g.,][]{yamamoto17}; the reaction of \ce{S^+} with \ce{H2} to form \ce{SH+} is endothermic, as is the reaction of \ce{SH+} (which can be formed by the reaction between atomic S and \ce{H3+}) with \ce{H2} to form SH$_2^+$.
\ce{H2S} has been speculated to be produced on grain surfaces by the sequential hydrogenation of atomic S on grain surfaces and subsequently released to the gas phase by thermal or nonthermal desorption processes, the latter of which should dominate the former at low temperatures ($\sim$10 K) \citep[e.g.,][]{garrod07}.
In the dark cloud L134N, the gas-phase \ce{H2S} abundance with respect to hydrogen nuclei is $4\times10^{-10}$ \citep{ohishi92}.
\ce{H2S} ice has not been detected in star-forming regions and the upper limit of the \ce{H2S}/\ce{H2O} ice abundance ratio is $\sim$1\% \citep{smith91}, corresponding to the \ce{H2S} ice abundance with respect to hydrogen nuclei of $\lesssim$10$^{-6}$.

Very little is known about \ce{PH3} in star-forming regions.
Neither gas-phase nor solid-phase \ce{PH3} has been detected in star-forming regions \citep[e.g.,][]{turner90,lefloch16}.
Like the gas-phase formation of \ce{H2S}, the formation of \ce{PH3} via gas-phase reactions is inefficient \citep{thorne84}.
Thus, the main formation pathway for gas-phase \ce{PH3} is the formation of \ce{PH3} ice by the sequential hydrogenation of atomic P on grain surfaces, followed by thermal or nonthermal desorption, as assumed in previous astrochemical models \citep[e.g.,][]{charnley94,aota12,chantzos20,sil21}.

To explore the effect of the chemical desorption of \ce{H2S} and \ce{PH3} on their gas-phase abundances, we conducted gas--ice astrochemical simulations using the modified rate-equation method \citep{garrod08} under dark-cloud physical conditions; the number density of hydrogen nuclei ($n_{\rm H}$), the temperature, and the visual extinction were set to $2\times10^4$ cm$^{-3}$, 10 K, and 10 mag, respectively.
The cosmic-ray ionization rate of \ce{H2} was set to $1.3\times10^{-17}$ s$^{-1}$.
In our rate-equation model, the chemistry is described by a three-phase model, where the gas phase, a surface of ice, and the chemically inert bulk ice mantle are considered \citep{hasegawa93}.
As nonthermal desorption processes, which are relevant to \ce{H2S} and \ce{PH3}, chemical desorption and photodesorption were considered.
The chemical desorption probability for reactions other than Reactions \ref{react:hs} and \ref{react:ph2} were calculated using the method of \citet{garrod07}, assuming $a=0.01$.
The photodesorption yields per incident far-ultraviolet (FUV) photons of \ce{H2S} and \ce{PH3} were set to 10$^{-3}$ \cite[see][]{fuente17}.
Additional details can be found in \citet{furuya15,furuya18}.

Although our kMC simulations have shown that the distribution of adsorption sites with different potential-energy depths plays a role in the chemical desorption of \ce{H2S} and \ce{PH3}, we must choose single values for the binding energy and the hopping energy of atomic H in the rate-equation model.
We set the hopping energy of atomic H to be 80 K, which corresponds to the low end of the distribution.
The $p_{\rm H_2S+H}$ was then $3\times10^{-3}$, and the $p_{\rm PH_3+H}$ was $3\times10^{-2}$.
The binding energy of atomic H was set to 300 K.

The amounts of elemental S and P available for gas and ice chemistry in the interstellar matter (ISM) are uncertain. 
In diffuse clouds, S is predominantly present in the gas phase, whereas P in the gas phase is depleted to some extent \citep{jenkins09}.
On the other hand, previous observations and modeling studies have suggested that the S and P abundances in star-forming regions are much lower than the values in diffuse clouds \citep[e.g.,][]{wakelam04,lefloch16}.
A large fraction of elemental S and P might be incorporated into refractory compounds during the evolution from diffuse clouds to denser clouds \citep[e.g.,][]{bergner19,cazaux21}.
Here, we assume that the elemental abundances of S and P available for gas and ice chemistry in the dark-cloud stage are lower than those observed in diffuse clouds by a factor 100 \citep{graedel82,wakelam08}.
The elemental abundances of S and P with respect to H were set to $1.5\times10^{-7}$ and $1.2\times10^{-9}$, respectively.
The elemental abundances of other elements were taken from \citet{aikawa99}.
Initially the species were assumed to be atoms or atomic ions except for hydrogen, which is in molecular form.

\begin{figure*}[ht!]
\plotone{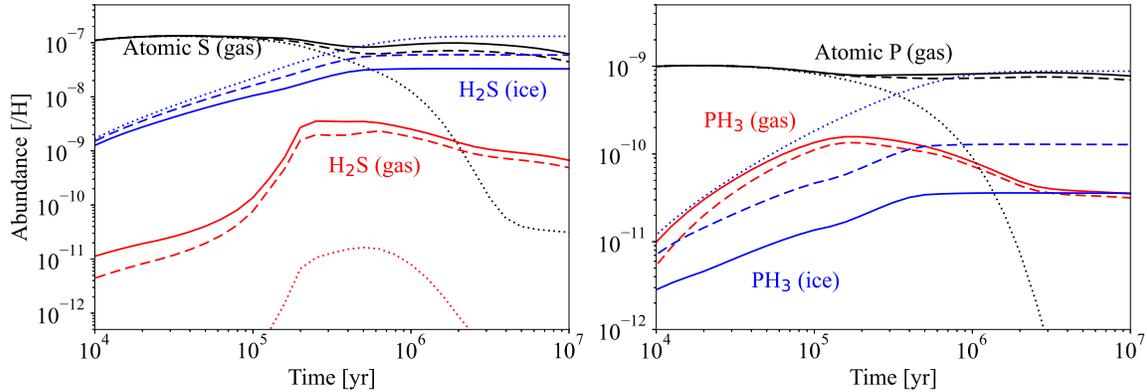}
\caption{Temporal evolution of the abundances of S-bearing species (left panel) and P-bearing species (right panel) with respect to H nuclei.
Solid lines represent models with $\pcds$ = 3 \% and $\pcdp$ = 4 \%, dashed lines represent models with $\pcds = \pcdp$ = 1 \%, and dotted lines represent models with $\pcds = \pcdp$ = 0 \% (i.e., without chemical desorption of \ce{H2S} and \ce{PH3}).
}
\label{fig:rateeq}
\end{figure*}

Figure \ref{fig:rateeq} shows the temporal evolution of S-bearing species (left panel) and P-bearing species (right panel), where the $\pcds$ and $\pcdp$ were varied.
In the model with $\pcds = \pcdp$ = 0\%, almost all the S and P are eventually confined in \ce{H2S} ice and \ce{PH3} ice, respectively.
The main production pathway for gas-phase \ce{H2S} and \ce{PH3} is photodesorption of the corresponding icy molecules by cosmic-ray-induced FUV photons.
In the model with $\pcds = 3$\% and $\pcdp = 4$\%, as constrained in the present work, the gas-phase abundances of \ce{H2S} and \ce{PH3} are higher by orders of magnitude than those in the model with $\pcd = 0$\%.
This result indicates that chemical desorption is the dominant dominant route for the supply of these gas-phase molecules.
Because of chemical desorption, the abundances of \ce{H2S} ice and \ce{PH3} ice are reduced compared with those in the model with $\pcd = 0$\%, and the dominant reservoirs of elemental S and P are the atomic forms.

The gas-phase \ce{H2S} abundance in the model with $\pcds = 3$\% is $(1-4)\times10^{-9}$ after 10$^5$ years.
The predicted value is similar to that observed in the molecular cloud TMC1-CP \citep[$(0.6-6)\times10^{-9}$, depending on the position in the cloud;][]{navarro-almaida20}.
The gas-phase \ce{PH3} abundance in the model with $\pcdp = 4$\% is $(3-20)\times10^{-11}$ after 10$^5$ years.
To the best of our knowledge, \ce{PH3} has not been detected in cold molecular clouds.
Future high-sensitivity observations of the \ce{PH3} $1_0-0_0$ transition at 266.9445136 GHz \citep{muller01,muller05} toward cold molecular clouds such as TMC-1 would provide an interesting test of the surface chemistry of P.

The model with $\pcd = 1$\%, which is often assumed in astrochemical models, predicts gas-phase abundances of \ce{H2S} and \ce{PH3} similar to those in the model with $\pcds = 3$\% and $\pcdp = 4$\%; the difference in the abundances is less than a factor of two.
This weak dependence indicates that the loops of \ce{H2S}--HS and \ce{PH3}--\ce{PH2} interconversions on dust grains are highly efficient and that the rate-limiting step of the desorption of \ce{H2S} and \ce{PH3} in our models is the adsorption of atomic H.
Notably, however, our rate-equation model, in which a single type of adsorption site is considered, may overestimate the rate of the loops of \ce{H2S}--HS and \ce{PH3}--\ce{PH2} interconversions.
In reality, deep-potential sites can trap atomic H, slowing the loops of \ce{H2S}--HS and \ce{PH3}--\ce{PH2} interconversions, as observed in our kMC models.
To accurately evaluate the effect of chemical desorption under the ISM conditions, the binding energy distribution for atomic H should be considered by solving a gas and ice chemical network.

\section{Summary}
\label{sec:concl}
Chemical desorption is caused by the energy released by reactions; after an exothermic surface reaction, some of the excess energy is dissipated into product's translational energy in the direction perpendicular to the surface, leading to desorption \citep[e.g.,][]{fredon17}.
Chemical desorption is usually included in modern gas--ice astrochemical models after \citet{garrod07};
however, its efficiency is poorly constrained, especially desorption from water ice.
\citet{oba19} experimentally studied chemical desorption upon the reaction of \ce{H2S} with H atoms on porous ASW.
\citet{nguyen21} conducted similar laboratory studies for the reactions of \ce{PH3} with H atoms.
These studies demonstrated that \ce{H2S} and \ce{PH3} can be lost from water ice surfaces by chemical desorption. 
They also estimated a chemical desorption probability for \ce{H2S} and \ce{PH3} {\it per reactive species} (i.e., per incident H atom) of $\sim$ 1\% on porous ASW.
As noted by \citet{oba18}, the desorption probability per incident H atom corresponds to the lower limit of the desorption probability {\it per reactive event}, which astrochemical models require as inputs, because a substantial fraction of adsorbed H atoms on ASW surfaces would be thermally desorbed and subsequently consumed by the \ce{H2} formation reaction on the surface.
In the present work, we constrained the desorption probability of \ce{H2S} and \ce{PH3} per reactive event on porous ASW by numerically simulating the laboratory experiments of \citet{oba19} and \citet{nguyen21}. 
We used kinetic Monte Carlo simulations in which the position and movement of each chemical species on surfaces were tracked over time and in which the binding energy distributions of atomic H and \ce{H2} were considered.
Our findings are summarized as follows.

\begin{enumerate}
{\item
The chemical desorption probability of \ce{H2S} and \ce{PH3} per hydrogenation event of the precursor species on porous ASW are constrained to $3 \pm 1.5$\% and $4 \pm 2$\%, respectively.
}

{\item
These probabilities are consistent with a theoretical model of chemical desorption proposed by \citet{fredon21}, with $\chi \sim 0.07$\% (where $\chi$ is a free parameter) describing the fraction of reaction energy transferred to the translational excitation of reaction products.
Whether the Fredon et al. model with $\chi \sim 0.07$ provides reasonable estimates of $\pcd$ for other reaction systems is unclear.
Experimental and numerical studies of additional systems are required for better understanding of the chemical desorption process.
}

{\item
As a byproduct, we constrained the binding energy distribution of atomic H.
The abundance of \ce{H2S} became $\sim$40\% and $\sim$70\% of the initial \ce{H2S} abundance in the 20 K and 30 K experiments by \citet{oba19}, respectively. 
These results indicate that $\sim$70\% of the adsorption sites on porous ASW should have $\edes({\rm H}) \lesssim 300$ K, whereas $\sim$40\% of adsorption sites should have $\edes({\rm H}) \lesssim 200$ K (see Section \ref{sec:specific}).
}
\end{enumerate}

Finally, we stress that, for the chemical reactions systems on which the present work is focused, basic chemical data are available in the literature (e.g., rate coefficients and binding energies).
Without these basic data provided by computational chemistry and experiments, we could not have constrained the chemical desorption probability. 
Basic chemical data are critical not only for astrochemical models of astrophysical objects but also for extracting chemical parameters from laboratory experiments of surface reactions. 

\acknowledgments
KF acknowledges Naoko Yokomura and Yuri Aikawa for their assistance in developing the kMC code used in this work.
YO acknowledges Naoki Watanabe, Akira Kouchi, Hiroshi Hidaka, and Nguyen Thanh for discussions of the experimental chemical desorption results.
We are grateful to the anonymous referees for providing valuable comments that helped improve the manuscript.
This work is partly supported by JSPS KAKENHI Grant numbers 17H06087, 20H05847, 21H01145, 21H04501, and 21K13967.
Numerical computations were carried out in part on a PC cluster at the Center for Computational Astrophysics, National Astronomical Observatory of Japan.

\begin{appendix}
\section{Additional model results}
\begin{figure}[ht!]
\plotone{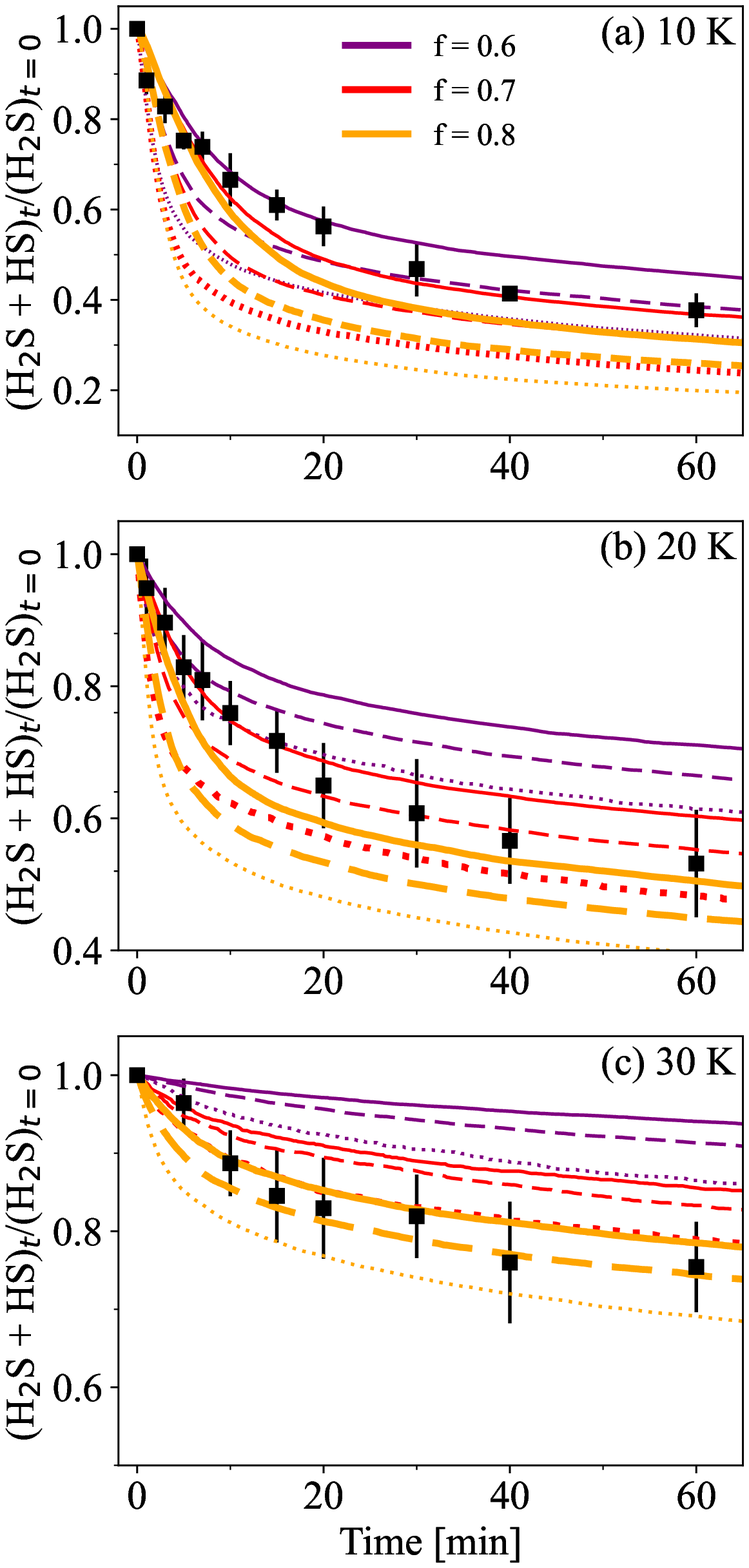}
\caption{Total sulfur (\ce{H2S} + HS) abundance with respect to the initial \ce{H2S} abundance on porous ASW at 10 K (panel a), 20 K (panel b), and 30 K (panel c) as functions of H-atom exposure time in the kMC simulations (lines) and in the experiments by \citet{oba19} (black squares).
Solid, dashed, and dotted lines represent the models with $\pcds$ values of 5\%, 10\%, and 20\%, respectively.
The models with different $f$ values are represented in different colors.
In all the models, $\edesmin({\rm H_2})$ is set to 150 K.
}
\label{fig:grid2}
\end{figure}

Figure \ref{fig:grid2} shows the results from additional models of the \ce{H2S} + H system, where $\edesmin({\rm H_2})$ is fixed at 150 K and $f$ and $\pcds$ are varied between 0.6 and 0.8 and between 5\% and 20\%, respectively.
Among the nine different models shown in the figure, only three models can reasonably well reproduce the 30 K experiments (shown by thick lines in the figure): the model with $f = 0.7$ and $\pcds = 20$\%, the model with $f=0.8$ and $\pcds = 5$\%, and the model with $f=0.8$ and $\pcds = 10$\%.
The first and third models overestimate the amounts of \ce{H2S} and HS desorbed from the surface at 10 K and 20 K compared with the amounts measured in the corresponding experiments.
The second model moderately well reproduces the experiments at 10 K, 20 K, and 30 K simultaneously; however, the model tends to overestimate the amount of \ce{H2S} and HS desorbed from the surface compared with the experiments at 10 K and 20 K.
Notably, this model is similar to our best-fit model ($f = 0.8$ vs 0.85 and $\pcds = 5$\% vs. 3\%); however, the reduced $\chi^2$ for this model is greater than that for the best-fit model (2.38 vs 1.85).

\epsscale{0.7}
\epsscale{1.0}

\section{Polynomial distribution vs. Gaussian distribution}
\begin{figure}[ht!]
\plotone{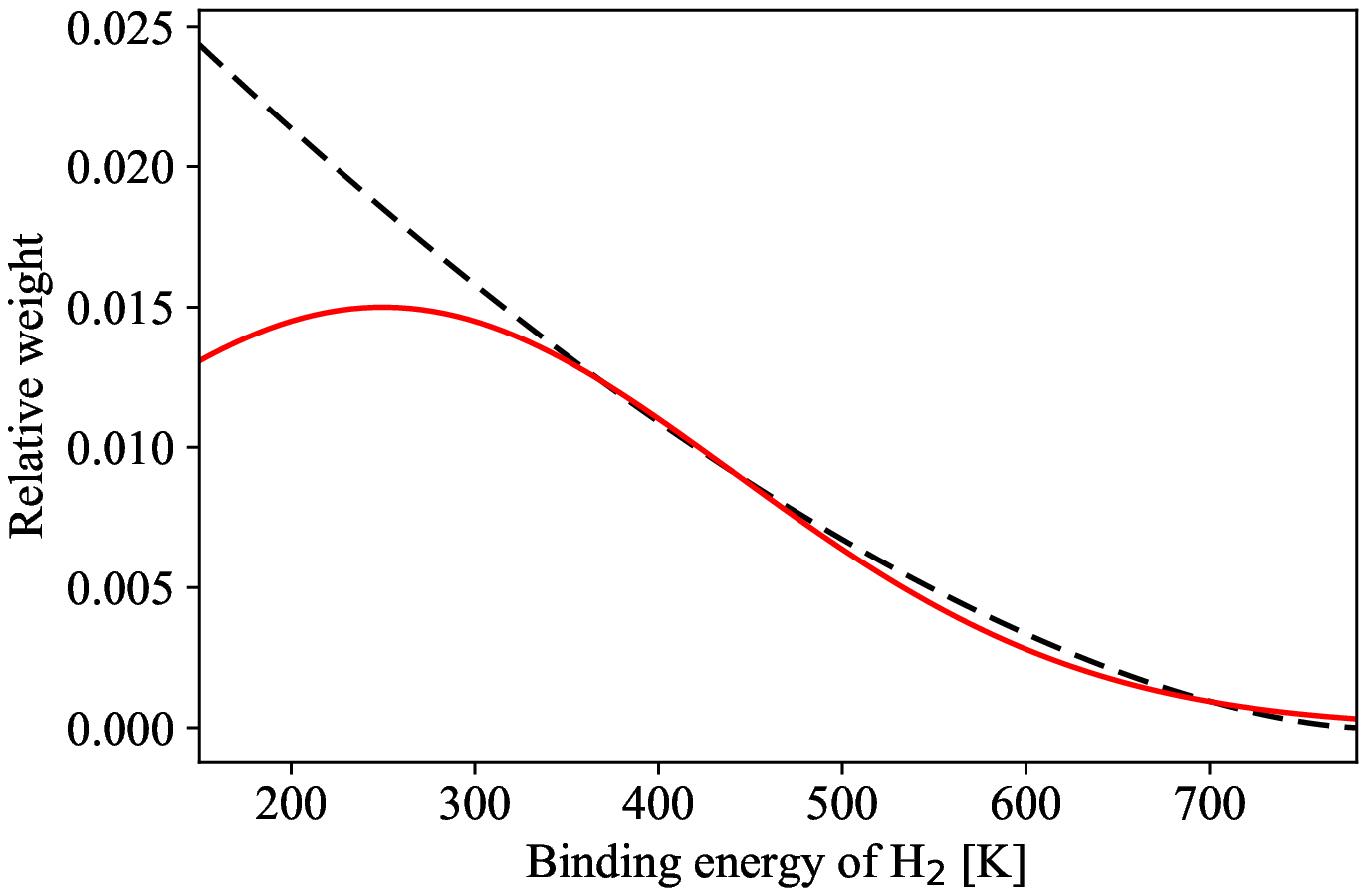}
\caption{Binding energy distribution for \ce{H2} assuming a Gaussian distribution (red) and a polynomial function (black). See the text for additional details.
}
\label{fig:gauss_dist}
\end{figure}

In our models, we assumed that the binding energy distributions for \ce{H2} are described by a polynomial function (Eq. \ref{eq:eb_dist}) from 780 K to $\edesmin$ on the basis of the experiments by \citet{amiaud06,amiaud15}.
The distribution at $\edes$ lower than $\sim$350 K was not constrained well in the experiments.
We simply extrapolated the polynomial function to lower $\edes$ values.
Here, we check the effect of this extrapolation on the modeling results. 
We tested some models in which the \ce{H2} binding energy followed a Gaussian distribution with a mean value of 270 K and a full-width at half-maximum (FWHM) of 190 K.
The lower boundary of $\edes({\rm H_2})$ was assumed to be 150 K, which roughly corresponds to the intermolecular potential energy of a \ce{H2}--\ce{H2O} dimer \citep[$\sim$120 K;][]{zhang91}.
As shown in Figure \ref{fig:gauss_dist}, this Gaussian distribution matches the polynomial distribution for $\edes \gtrsim 350$ K; by contrast, for lower $\edes$, the two distributions deviate.

\begin{figure}[ht!]
\plotone{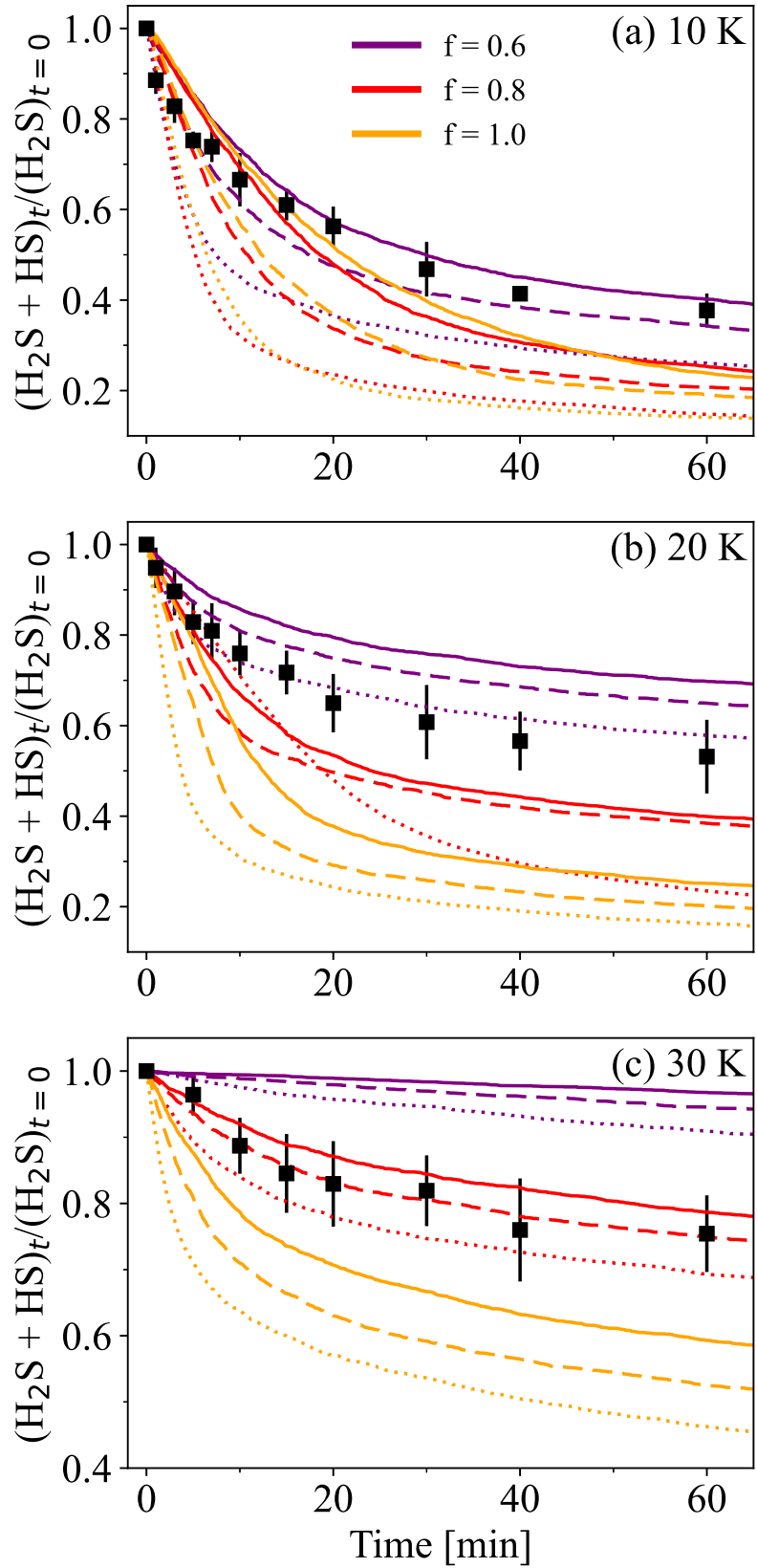}
\caption{Total sulfur (\ce{H2S} + HS) abundance with respect to the initial \ce{H2S} abundance on porous ASW at 10 K (panel a), 20 K (panel b), and 30 K (panel c) as functions of H-atom exposure time in the models, where the binding energies for atomic H and \ce{H2} follow a Gaussian distribution (lines).
Black squares represent the experiments by \citet{oba19}.
Solid, dashed, and dotted lines represent the models with $\pcds$ of 3\%, 5\%, and 10\%, respectively.
The models with different $f$ are represented in different colors.
}
\label{fig:gaussian}
\end{figure}

Figure \ref{fig:gaussian} shows the results of a small grid of models, where $f$ and $\pcds$ are varied between 0.6 and 0.8, and between 3\% and 10\%, respectively.
As shown, no model can reproduce the experiments at 10 K, 20 K, and 30 K simultaneously.
We conclude that there is no clear advantage to using the Gaussian distribution instead of the polynomial distribution.

\epsscale{0.8}
\epsscale{1.0}

\epsscale{0.7}
\epsscale{1.0}

\end{appendix}

\bibliography{sample63}{}
\bibliographystyle{aasjournal}





\end{document}